\documentclass[10pt,onecolumn,draft]{IEEEtran}

\usepackage{amsmath}
\usepackage{amsthm}
\usepackage{amssymb}
\usepackage{graphicx}
\usepackage{rotating}
\usepackage{flushend}
\usepackage{verbatim}
\allowdisplaybreaks
\usepackage{mathtools}
\usepackage{tikz}

\newsavebox{\tempbox}

\usepackage{xcolor}

\newcommand{\logt}{\log_{2}}
\newcommand{\dx}{\mathrm{d}}

\newcommand{\abs}[1]{\left| #1 \right| }

\newcommand{\set}[1]{\left\{ #1 \right\}}

\newcommand{\defn}{\triangleq}

\newcommand{\mbf}[1]{\mathbf{#1}}

\newcommand{\mcf}[1]{\mathcal{#1}}
\newcommand{\idc}[1]{1\left\{#1\right\} }

\makeatletter
\newcommand{\subalign}[1]{%
  \vcenter{%
    \Let@ \restore@math@cr \default@tag
    \baselineskip\fontdimen10 \scriptfont\tw@
    \advance\baselineskip\fontdimen12 \scriptfont\tw@
    \lineskip\thr@@\fontdimen8 \scriptfont\thr@@
    \lineskiplimit\lineskip
    \ialign{\hfil$\m@th\scriptstyle##$&$\m@th\scriptstyle{}##$\crcr
      #1\crcr
    }%
  }
}
\makeatother

\DeclareMathOperator{\markov}{\setlength{\unitlength}{.5cm} \begin{picture}(1,1)  \put(0,.22){\line(1,0){1}}  \put(.5,.22){\circle{.3}}   \end{picture}}

\DeclareMathAlphabet\mbcf{OMS}{cmsy}{b}{n}

\setkeys{Gin}{draft=false}

\newtheorem{theorem}{Theorem}
\newtheorem{define}[theorem]{Definition}
\newtheorem{lemma}[theorem]{Lemma}
\newtheorem{cor}[theorem]{Corollary}
\newenvironment{remark}[1][Remark]{\begin{trivlist}
\item[\hskip \labelsep {\bfseries #1}]}{\end{trivlist}}

\IEEEoverridecommandlockouts

\begin{document}
\title{Secret key authentication capacity region, Part \textrm{II}: typical authentication rate}
\author{Eric Graves, Jake Perazzone, Paul Yu, and Rick Blum
\thanks{This material is based upon work partially supported by the U. S. Army Research Laboratory and the U. S. Army Research Office under grant number W911NF-17-1-0331 and by the National Science Foundation under grants ECCS-1744129 and CNS-1702555.}
\thanks{Eric Graves and Paul Yu are with the Army Research Lab,  Adelphi, MD
  20783, U.S.A. \texttt{\{eric.s.graves9, paul.l.yu\}.civ@mail.mil } }%
\thanks{Jake Perazzone and Rick Blum are with the Department of Electrical and Computer Engineering,
 Lehigh University,
 Bethlehem, PA 18015, U.S.A. 
\texttt{\{jbp215,rb0f\}@lehigh.edu}}%
}
\maketitle

\begin{abstract}
This paper investigates the secret key authentication capacity region. 
Specifically, the focus is on a model where a source must transmit information over an adversary controlled channel where the adversary, prior to the source's transmission, decides whether or not to replace the destination's observation with an arbitrary one of their choosing (done in hopes of having the destination accept a false message). 
To combat the adversary, the source and destination share a secret key which they may use to guarantee authenticated communications.
The secret key authentication capacity region here is then defined as the region of jointly achievable message rate, authentication rate, and key consumption rate (i.e., how many bits of secret key are needed).

This is the second of a two part study, with the studies separated by how the authentication rate is measured. 
Here, the authentication rate is measured by the minimum of the maximum probability of false acceptance where the minimization is over all highly probable subsets of observations at the adversary.
That is, consider the maximum probability of false authentication as a function of the adversary's observation, and the adversary's observation as a random variable.
The authentication rate is then measured as the smallest number for which the probability that the maximum probability of false authentication is greater than said number is arbitrary small.
This is termed typical authentication, since it only needs to consider adversarial observations which are typical with the transmission. 
Under this measure of authentication matching inner and outer bounds are determined.
Not surprisingly, the region can be expressed in terms of classical measures on the channel's information.
Of importance, the authentication rate region expressed shows that there is a trade-off between message rate and authentication rate; more specifically, the two must share the channel's capacity.

\end{abstract}

\section{Introduction}

Authentication is inherently a physical layer problem; any protocol that labels data as valid or invalid naturally creates a bifurcation of the physical layer observations. 
What's more, this labeling should degrade the performance of the communication system in comparison to a system which does not require authentication, since any possible observation which is labeled as inauthentic can no longer contribute to the probability of reliably decoding.
Our goal with this series of papers is to generally explore these trade-offs.
In particular, we wish to explore this trade-off in a model previously considered by Lai et al.~\cite{lai2009authentication} and as a sub-case by Gungor and Koksal~\cite{gungor2016basic}.
With this particular model, information must be sent in the presence of an adversary. 
This adversary is particularly powerful in that it can observe a noisy version of the transmitted data, and then arbitrarily decide the destination's observation.
On the other end, the communicating parties are allowed to share a secret key prior to communications.
For this model, our goal is to derive a classical information theoretic ``rate region'' that describes the trade-off between message rate, authentication rate, and the amount of secret key required (termed the key consumption rate).

This work has been split into two papers since, in the course of our efforts to obtain the desired rate region characterization, we discovered that the traditional metric for authentication (the maximum probability of false authentication) does not necessarily represent the true strength of the systems' authentication capability.
Indeed, the traditional metric is beholden to extremely unlikely events occurring in the communication channel; for example, a noisy binary symmetric channel acting as a noiseless channel.
As a result, designing codes around the traditional metric leads to codes which are designed with extremely unlikely cases in mind.
Upon this discovery, we formulated a new metric which only considers ``typical'' behaviour of the communication channel, with all other behavior being written off as loss. 
With Part I, we explored this trade-off under the traditional metric. 
Here in Part II, we explore this trade-off with our new metric and characterize the trade-off region. 
The need to split the papers based upon choice of metric is done primarily to allow flexibility in how the results are presented, with the traditional metric's dependence on unlikely empirical channels dictating a notation where the various information theoretic terms are functions of probability distributions, while the new metric allows for a (in our opinion) simpler presentation where the information theoretic terms are functions of random variables.
Additionally, while the focus for part I is on achievability, here we must also focus on the converse.

Authentication is an important topic considered outside of the information theoretic literature. 
Some examples include: Yu et al.~\cite{Yu08} who used spread spectrum techniques in addition to a covert channel to ensure authentication, 
Xiao et al.~\cite{Xiao08} who used the unique scattering of individual users in indoor environments to authenticate packets, and Korzhik et al.~\cite{Korzhik07} who make use of a (possibly noisy) initialization setup to create unique correlations which then allow for detection. 
These methods, while perhaps more suitable for application, use tools that are insufficient in determining the various information theoretic measures considered here.
Instead, what they highlight is a concern for authentication that should not be ignored. 
With this work, we hope to provide insight into the general problem, and provide baselines to what is possible.

On the other hand, authentication has only somewhat been considered from the information theoretic viewpoint.
In particular, it can be argued that Blackwell et al.~\cite{blackwell1960capacities} and their study of the  \emph{arbitrarily varying channel} (AVC) was the first true study of authentication. 
For the AVC, an adversary can at will choose the state of the communication channel between the two communicating parties.
This classic work and those that followed, such as \cite{wolfowitz2012coding,dobrushin1975coding,csiszar1988capacity}, all considered the maximum communication rate that can be obtained subject to an arbitrarily small probability of error (over any choice of communication states by the adversary). 
Note, this indeed implies that a decoded message would be authentic because the probability of error must take into account the adversaries actions.
In this vein, Ahlswede~\cite{ahlswede1978elimination} considered the communication rate over an AVC when the source and destination share a secret key. 
More specifically, Ahlswede gave the two communicating parties access to shared randomness, which must be kept private from the adversary prior to transmission.
For Ahlswede, allowing this secret key dramatically improved the communication rate, essentially transforming AVCs into a compound channel.

While these papers do examine an aspect of authentication, one can also argue that they are much too strict in their operational requirement. 
Today, the detection of the adversaries involvement is a strong enough result for many fields of security; for example, in quantum key distribution a system is considered operational even though the adversary can reduce the key rate to zero by measuring the data.
In our case, it makes even more practical sense to forgo such a harsh operational requirement. 
That is, if an adversary wanted to reduce the communication rate to zero between two parties in practice, they would simply need a strong enough jammer. 
Of course, simply jamming a signal is different than trying to have a node accept a fabricated message as authentic. 
This is the stance we adopt here: when the adversary is attacking, a system is operational if it can decode the correct message \emph{or} detect the attack; when the adversary is not attacking, we want the system to communicate as much data as possible.

Adopting this viewpoint, works by Jiang~\cite{jiang2014keyless,jiang2015optimality}, Graves et al.~\cite{graves2016keyless}, Kosut and Kliewer~\cite{kosut2018authentication}, and Beemer et al. \cite{beemer2019authentication} all consider authentication over an AVC without a secret key.
In particular, Jiang considered the sub-case of AVC where the output of the AVC was independent of the legitimate parties input for all but a single channel state. 
Graves et al. considered a general AVC where the adversary is given the side information of which message is being transmitted, while Kosut and Kliewer considered the general AVC case.
Finally, Beemer et al. considered a binary AVC, where the adversary is allowed to observe the source's transmission through a noisy channel before choosing the channel state.  
Each of these works avoids looking at the strength of the authentication capability, and instead only considers the data rate given the maximum probability of false authentication goes to zero.

Works considering secret key-based authentication have their genesis in Simmons~\cite{Auth}, who considered a special case of the model presented here where all channels are noiseless. 
The fundamental distinction separating the problems of keyless and secret key authentication is that the former relies on exploiting nature of the communication channels, while the latter relies on exploiting a finite resource.
Later came the works of Lai et al.~\cite{lai2009authentication} and that of  Gungor and Koksal~\cite{gungor2016basic}, who both consider generalizations of Simmons' model with noisy channels. 
Each of these works has aspects which could be strengthened.
Lai et al. require the amount of secret key bits to be asymptotically negligible when compared with the blocklength of the transmission.
In doing so, though, they can make no distinction in the importance of verifying ten versus ten thousand bits of data. 
Meanwhile, Gungor and Koksal's coding scheme is inefficient and mismanages the key by unnecessarily using it in a way that favors the adversary.
Furthermore, their work does not attempt\footnote{Although we did endeavour to extract such a rate region from their works, we were unable to do so and instead had to settle for an outer bound. Regardless, our results improve on an outer bound to their inner bound. See Part 1 of these works.} to explicitly derive such a region, instead opting for a presentation of error exponents. 

Once again, for this paper, we look to characterize the trade-off between information rate, strength of authentication, and the amount of required key.
In this setting, with our new typical authentication rate metric, we are able to derive a matching inner and outer bound, thus completely characterizing the region.
For the inner bound, we will use a coding scheme similar to that of Part I.
On the other hand, for the converse, we use results from Graves and Wong~\cite{graves2018inducing} which allow us to directly turn the operational requirement of authenticity into bounds on mutual information terms.

We conclude this introduction by presenting the notation that will be used throughout the paper in Section~\ref{sec:notation}.
Following this, we shall present the exact channel model, and its relevant definitions in Section~\ref{sec:model}.
Section~\ref{sec:prev_work}, then, revisits past work on this model, describing the works of~\cite{lai2009authentication},~\cite{Auth}, and~\cite{graves2018inducing} in more detail as we believe understanding the past schemes will allow for a better understanding of our approach.
We will also describe the results of Csisz{\'a}r and K{\"o}rner~\cite{csiszar1978broadcast} on the discrete memoryless broadcast channel with confidential communications, which will provide the basis for our direct scheme.
Fundamental results are then presented in Section~\ref{sec:main_cont}, and examples of given in Section~\ref{sec:examples}.
Proofs can be found in appendices.


\subsection{Notation}\label{sec:notation}

Uppercase letters will be used to denote \emph{random variables} (RVs) and lowercase letters will be used to denote constants. 
The probability of event $\mcf{A}$ is denoted $\Pr (\mcf{A} )$. 
Function $p$ with subscript RV will be used to denote the probability distribution over the RV (i.e., $p_{X}(x) = \Pr (X=x)$). 
To simplify presentation, the subscript may be suppressed when clear. 
Calligraphic font or curly brackets will be used to denote sets, for instance $\mcf{Y} = \set{1,\dots, 10}$. 
The only exceptions to this are the set of positive real numbers, denoted $\mathbb{R}^+$, and the set of positive integers, denoted $\mathbb{Z}^+$. 
Subscripts will generally be used for bookkeeping purposes. 
While $|$ denotes the word ``given,'' and $:$ ``subject to.''

The function $\times$ will be used to denote the Cartesian product.
We will frequently need to use the Cartesian product of $n$ (where $n$ will denote the block length of a given code) correlated RVs, constants, and sets.
This need arises so frequently that we denote these Cartesian products by bold face.
For instance, $\mbf{X} = \times_{i=1}^n X_i = (X_1,\dots,X_n)$ and $\mbcf{X} = \times_{i=1}^n \mcf{X}.$
When using this notation with a probability distribution the terms in the product are uncorrelated. 
For example, given a probability distribution $p_{X}$ over $\mcf{X}$
\[
\mbf{p}_{X}(\mbf{x}) = \prod_{i=1}^n p_{X}(x_i)
\]
for each $\mbf{x} \in \mbcf{X}$.

The indicator function of an event $\mcf{A}$ is denoted $\idc{\mcf{A}}$, that is $\idc{\mcf{A}} = 1$ if $\mcf{A}$ occurs, otherwise $\idc{\mcf{A}} = 0$.

The set of all probability distributions on a certain set, say $\mcf{X}$, is denoted by $\mcf{P}(\mcf{X})$, likewise $\mcf{P}(\mcf{Y}|\mcf{X})$ denotes the probability distributions of $\mcf{Y}$ conditioned on elements of $\mcf{X}$. 
The set $\mcf{P}(\mcf{Y}\gg \mcf{X})$ represents a special subset of $\mcf{P}(\mcf{Y}|\mcf{X})$, where for each $v \in \mcf{P}(\mcf{Y} \gg \mcf{X}) $ and $y \in \mcf{Y}$ there exists at most one $x \in \mcf{X}$ such that $v(y|x) > 0$. 
Note, for random variables $X,Y,Z$, if $p_{Y|X} \in \mcf{P}(\mcf{Y}\gg \mcf{X})$, then $X,~Y,~Z$ form a Markov chain, $X \markov Y \markov Z$.

Another special subset of the distributions is the possible ``empirical distributions'' (or type classes) for a given $n$-length sequence, denoted $\mcf{P}_{n}(\cdot)$. 
The empirical distribution of sequence $\mbf{x}$, denoted $p_{\mbf{x}}$, is the distribution defined by the proportion of occurrences of $x$ in sequence $\mbf{x}$.
In other words, 
\[ p_{\mbf{x}}(b) \defn  \frac{\sum_{i=1}^n \idc{x_i = b} }{n}, \quad\quad \forall b \in \mcf{X}.
\]
This follows similarly for empirical conditional distributions, but we further list the empirical distribution of the conditioning value, such as $\mcf{P}_n(\mcf{Y}|\mcf{X};\rho)$ for $\rho\in \mcf{P}_n(\mcf{X})$. 
Here, the empirical conditional distribution of $\mbf{y}$ given $\mbf{x}$ is defined by 
\[ p_{\mbf{y}|\mbf{x}}(b|a) \defn  \frac{\sum_{i=1}^n \idc{y_i =b} \idc{x_i = a} }{\sum_{i=1}^n \idc{x_i =a }}, \quad\quad \forall a\times b \in \mcf{X} \times \mcf{Y} .
\]
For each $\mu \in \mcf{P}_{n}(\mcf{Y}|\mcf{X};\rho)$ and $\rho \in \mcf{P}_{n}(\mcf{X})$, the type class of $\mu$ given a $\mbf{x}$ such that $p_{\mbf{x}} = \rho$ is denoted
\[
\mbcf{T}_{\mu}(\mbf{x}) \defn \set{\mbf{y} : p_{\mbf{y}|\mbf{x}} = \mu }.
\]

Black board bold (other than the two exceptions discussed earlier) is used to denote functions which are averaged over RVs.
Of particular importance is $\mathbb{E}$ which denotes the expectation operator.
Other important functions are entropy and mutual information denoted (respectively) by
\begin{align*}
\mathbb{H}(Y|X) &= -\sum_{\subalign{y &\in \mcf{Y},\\ x & \in \mcf{X}}} p(y,x) \logt p(y,x), \\
\mathbb{I}(Y;X|U) &= \sum_{\subalign{y&\in \mcf{Y},\\x &\in \mcf{X},\\ u&\in \mcf{U}}} p(y,x,u) \logt  \frac{p(y,x|u)}{p(y|u)p(x|u)}\\
&= \mathbb{H}(Y|U) + \mathbb{H}(X|U) - \mathbb{H}(X,Y|U) ,
\end{align*}
for discrete random variables $X$, $Y$, and $U$.

In addition to the traditional absolute value, for any $a \in \mathbb{Z}$, $\mbf{a} \in \mathbb{R}^n$, and set $\mcf{A}\subseteq \mcf{X}$ define the following:
\begin{align*}
\abs{\mbf{a}} &=\sum_{i=1}^n |a_i|  \\
\abs{a}^+ &= a \idc{a > 0} \\
\abs{a}^- &= a \idc{a < 0} \\
\abs{\mcf{A}} &= \sum_{x \in \mcf{X}} \idc{x \in \mcf{A}}.
\end{align*}

Finally, the $O$ function from the Bachmann-Landau notation will be employed here. 
That is, by writing $g(x,n) = f(x,O(h(n)))$, we are saying that there exists a constant such $\zeta$, independent of $n$, such that
$$ |g(x)|  \leq \max_{ r \in [-\zeta h(n),\zeta h(n)]} f(x,r).$$

\section{Model}\label{sec:model}

\tikzstyle{block} = [draw, fill=white, rectangle, 
    minimum height=20pt, minimum width=20pt, text centered]

\begin{figure}
 \begin{center}
\begin{tikzpicture}
\node[block] (enc) at (0,0) {$\begin{array}{c} \text{Alice} \\ f \end{array}$};
\node[block] (chan) at (3.2,0) {$\begin{array}{c} \text{Channel} \\ p_{Y|X} \end{array}$};
\node[block] (chanz) at (1,-3) {$\begin{array}{c} \text{Channel} \\ p_{Z|X} \end{array}$};
\node[block] (dec) at (8,-1.5){$\begin{array}{c} \text{Bob} \\ \varphi  \end{array}$};
\node[block] (adver) at (3.2,-3){$\begin{array}{c} \text{Gr{\'i}ma} \\ \psi \end{array}$};
\node[circle] (switch) at (6,-1.5)  {$\begin{array}{c}~  \end{array}$ };
\draw[->,thick] (-1.5,0) -- node[above]{$M$} (-1.5,0)  -- (enc.west) ;
\draw[->,thick] (enc.east)  node[above right]{$\mbf{X}$}  -- (chan.west) ;
\draw[->,thick] (1,0) -- (chanz.north) ;
\draw[->,thick] (chan.east) -- (6,0) node[right]{$\mbf{Y}$} -- (switch.north)  ;
\draw[->,thick] (adver.east) -- (6,-3) node[right]{$\mbf{\hat Y}$} -- (switch.south) ;
\draw[-,dotted,thick] (switch.south) -- (switch.east);

\draw[->,thick] (dec.east) node[above right]{$\hat M$} -- (9.5,-1.5) ;
\draw[->,thick] (chanz.east) node[above right]{$\mbf{Z}$}  --(adver.west);
\draw[<->,thick,dashed] (enc.north) -- (0,1.2) -- (4,1.2) node[above]{$K$} -- (8,1.2)-- (dec.north) ;
\draw[->,thick,dashed] (adver.north) -- (3.2,-1.5) 
-- 
(switch.east) ;
\draw[->,thick] (switch.east) node[above right] {$\mbf{\hat Y}$}  -- (dec.west);
\end{tikzpicture}
 \caption{Channel model where Gr{\'i}ma has chosen to interlope.}
\label{fig:II}
\end{center}
\end{figure}
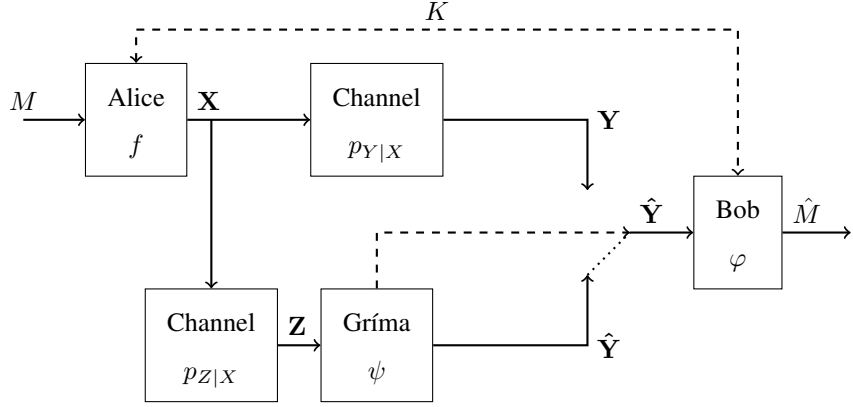

In this communication model (pictured in Figure~\ref{fig:II}), Alice wishes to send a message $M$, uniformly distributed on $\mcf{M} = \{1,\dots,2^{nr}\}$ with $n \in \mathbb{Z}^+$  and $r \in\mathbb{R}^+$, to Bob over a (to be defined later) \emph{discrete memoryless-adversarial interlope channel$(p_{Y|X},p_{Z|X})$} (DM-AIC$(p_{Y|X},p_{Z|X})$).
The DM-AIC$(p_{Y|X},p_{Z|X})$ is a channel controlled by Gr{\'i}ma\footnote{Chosen for Gr{\'i}ma Wormtongue from \emph{Lord of the Rings} by J.R.R. Tolkien. Gr{\'i}ma was an advisor to the King of Rohan, while secretly an agent of Saruman. Thus his role was to listen to information presented to the King and manipulate it towards Saruman's agenda. This seemed more appropriate than ``Eve,'' since the adversary does not only take the role of eavesdropper.}, whose objective is to get Bob to accept a false message.
To assist Alice and Bob, prior to the communication, Alice and Bob share a secret key, $K$, chosen uniformly over $\mcf{K} \defn \set{1, \dots, 2^{n  \kappa}}$, where $\kappa \in \mathbb{R}^+$.
For simplicity, we assume that $2^{nr}\in \mathbb{Z}^+ $ and $2^{n\kappa}\in \mathbb{Z}^+$.
To transmit this message, Alice uses an \emph{encoder} that selects an $n$-symbol channel input sequence $\mbf{X}$ as a (possibly stochastic) function of the message $M$ and key $K$.
Throughout this paper, $f \in \mcf{P}(\mbcf{X}|\mcf{M} ,\mcf{K})$ will be used to denote the stochastic relationship between the encoder's channel input sequence given the message and secret key.

On the other end, Bob uses a \emph{decoder} to estimate the message as a function of the channel's output sequence, either $\mbf{Y}$ or $\mbf{\hat Y}$, and the shared key. 
The ``$\mbf{!}$'' symbol is to be representative of the decoder declaring their observation is inauthentic. 
Similar to the encoder, the decoder will be identified by a conditional probability distribution $\varphi \in \mcf{P}(\mcf{M} \cup \{ \mbf{!} \} |\mbcf{Y},\mcf{K})$.

We now return to the discrete memoryless-adversarial interlope channel$(p_{Y|X},p_{Z|X})$. 
If Gr{\'i}ma chooses his own sequence for the channel to output, it will be called interloping, and $\mbf{\hat Y}$ will denote the channel's output sequence.
When interloping, Gr{\'i}ma may arbitrarily choose the value of $\mbf{\hat Y}$ as a function of his own observation $\mbf{Z}$, where the probability that $\mbf{Z}=\mbf{z}|\{\mbf{X} = \mbf{x}\}$ is $\mbf{p}_{Z|X}(\mbf{z}|\mbf{x})  = \prod_{i=1}^n p_{Z|X}(z_i|x_i)$.
Thus, when Gr{\'i}ma interlopes, we shall make the assumption that the probability $\mbf{\hat Y}=\mbf{\hat y}|\{ \mbf{Z} = \mbf{z} \}$ is $\psi(\mbf{y} |  \mbf{z})$ for some $\psi \in \mcf{P}(\mbcf{Y}|\mbcf{Z})$.
In general, it should be assumed that $\psi \in \mcf{P}(\mbcf{Y}|\mbcf{Z})$ will be chosen to minimize the authentication measure (to be discussed more later).
On the other hand, when Gr{\'i}ma does not interlope, $\mbf{Y}$ will denote the channel's output sequence, where specifically the probability that $\mbf{\hat Y}=\mbf{y}|\{\mbf{X} = \mbf{x}\}$ is $\mbf{p}_{Y|X}(\mbf{y}|\mbf{x})  = \prod_{i=1}^n p_{Y|X}(y_i|x_i).$
Note this channel is not a true memoryless channel since Gr{\'i}ma does not need to act in a memoryless fashion on the symbols. 
Instead, the $p_{Y|X}$ and $p_{Z|X}$ in a DM-AIC$(p_{Y|X},p_{Z|X})$ only specify the memoryless channels that connect Alice to Bob (if Gr{\'i}ma does not interlope) and Alice to Gr{\'i}ma, respectively.

\subsection{Operational Definitions}

For this part of the study, we adopt the \emph{typical authentication rate}, for a given \emph{authentication failure tolerance}.
\begin{define}
A code $(f,\varphi)$ has \emph{typical authentication rate} $\alpha$ and \emph{authentication failure tolerance} $\epsilon$ if 
$$ \sup \left\{ a \in \mathbb{R}^+ :  \sup_{\psi \in \mcf{P}(\mbcf{Y}|\mbcf{Z})} \Pr \left(  -n^{-1} \logt \omega_{f,\varphi}(\mbf{Z},M,K) < a \right) \leq \epsilon \right\}  \leq \alpha $$ 
where 
$$
\omega_{f,\varphi}(\mbf{z},m,k) \defn \sum_{\mbf{y}} \psi(\mbf{y}|\mbf{z}) \sum_{m' \in \mcf{M} -\{m\}}\varphi(m' |\mbf{y},k ).
$$
\end{define}
Recall from Part I of this study, $\omega_{f,\varphi}(\mbf{z},m,k)$ can be interpreted as the probability of false authentication given Gr{\'i}ma observed $\mbf{z},$ the true message is $m$, and the key is $k$.
In that sense, a code with a typical authentication rate of $\alpha$ with an authentication failure tolerance of $\epsilon$ guarantees that, with probability at least $1-\epsilon$, Gr{\'i}ma's observed sequence only provides Gr{\'i}ma with a $2^{-n\alpha}$ probability to fool Bob into authentication. 
In comparison, average authentication is the probability of false authentication, averaged over all possible observations by Gr{\'i}ma. 
Thus, the typical authentication rate is a measure of Gr{\'i}ma's ability to falsely inject a message, as long as the channel behaves as expected.

Moreover, the authentication failure tolerance acts as a measure of channel deviation in the same way as the final operational definition, the \emph{probability of message error}. 
\begin{define} \label{def:dm-bccc}
The \emph{probability of message error} is
\begin{equation*}
\varepsilon_{f,\varphi} \defn   \mathbb{E}\left[ \varepsilon_{f,\varphi}(M,K) \right]
\end{equation*}
where
\begin{align*}
\varepsilon_{f,\varphi}(m,k) &\defn \mathbb{E} \left[  1 - \varphi(m|\mbf{Y},k) \right] \\
&=1- \sum_{\mbf{x},\mbf{y}} \varphi(m|\mbf{y},\mbf{k}) \mbf{p}_{Y|X}(\mbf{y}|\mbf{x}) f(\mbf{x}|m,k).
\end{align*}
\end{define}
\begin{remark}
Neither the authentication failure tolerance or the probability of message error is measured per transmitted symbol.
\end{remark}

Combining these operational parameters, we define the following code measure. 
\begin{define}\label{def:average code}
Code $(f,\varphi)$ is a $(r,\alpha,\kappa,\delta,n)$-\emph{typical authentication} (TA) code for DM-AIC$(p_{Y|X},p_{Z|X})$ if it has blocklength $n$, message rate at least $r$, typical authentication rate at least $\alpha,$ key consumption rate at most $\kappa$, and both probability of message error and authentication failure tolerance less than $\delta$.
\end{define}

The goal of this work is to determine the inherent trade-offs between the message rate, authentication rate, and key consumption rate, when the probability of message error and authentication failure tolerance go to zero. 
Note that requiring the authentication failure tolerance go to zero yields codes for which almost surely the probability of false authentication will have an upper bound of $\exp(-(n)\text{ authentication rate})$.
As with the average authentication capacity region, the typical authentication capacity region is defined as a limit point of the operational measures as the blocklength goes to infinity. 
\begin{define}
A triple $(r,\alpha,\kappa)\in \mcf{C}_{\mathrm{TA}}(p_{Y|X},p_{Z|X})$ if $r>0$ and for each $i \in \mathbb{Z}^+$ there exists a $(r_i,\alpha_i,\kappa_i,\delta,i)$-TA code for DM-AIC$(p_{Y|X},p_{Z|X})$ where 
\[
\lim_{i \rightarrow \infty} \abs{ (r_i,\alpha_i,\kappa_i,\delta_i) - (r,\alpha,\kappa,0)} = 0.
\] 
The set $\mcf{C}_{\mathrm{TA}}(p_{Y|X},p_{Z|X})$ is called the \emph{typical authentication capacity region}.
\end{define}
\begin{remark}
This region is, once again, closed (but not necessarily convex) by definition.
\end{remark}

\begin{remark}
The values of $r=0$ are excluded from this region, since these codes transmit asymptotically $0$ information per symbol.
Furthermore, eliminating the case where $r=0$ yields a $\mcf{C}_{\mathrm{TA}}(p_{Y|X},p_{Z|X})$ that is convex. 
If, instead, we were to allow $r=0$, then the new value $\mcf{C}_{\mathrm{TA}}(p_{Y|X},p_{Z|X})$ would simply be the union of the region presented in our results, and the set of $(0,a,b)$ for all non-negative real numbers $a$ and $b$. 
Indeed, consider the case where there is only a single message to transmit. 
In this case it would be impossible for Gr{\'i}ma to replace it with an alternative, and thus provide an infinite authentication rate, without consuming any key.  
Therefore, the case of $r=0$ is not really interesting in our context. 

\end{remark}

\section{Background}\label{sec:prev_work}

For the reader's convenience, we shall briefly describe the coding schemes of Lai et al.~\cite{lai2009authentication} and Simmons~\cite{Auth}. 
While our coding scheme is novel in the sense that it has not previously appeared, it does share a design philosophy with Lai et al. and with Simmons. 
These schemes separate in an intuitive way, with Lai et al.'s scheme exploiting the channel for authenticity and Simmon's scheme exploiting only the secret key for authenticity. 
In addition to these coding schemes, it will be helpful to further discuss codes for the \emph{discrete broadcast channel with confidential communications} (DM-BCCC), from~\cite[Theorem~17.13]{CK}.
Codes for the DM-BCCC will act as a base code to with which we can use Lai's strategy (although not his direct coding scheme). 

For the converse, it will be helpful to briefly discuss information stabilizing random variables, introduced by Graves and Wong~\cite{graves2018inducing}.
These constructed random variables will provide us with a method by which to turn the operational requirements for authentication into requirements on information terms similar to how Fano's inequality turns the requirement for a small probability of error into a requirement on a conditional entropy.

\subsection{Lai's Strategy} \label{sec:back:lai}

Lai et al.~\cite{lai2009authentication} used the strategy of having Alice explicitly send the value of the secret key to Bob, while simultaneously obfuscating this value from Gr{\'i}ma. 
Since it will be of direct use here, we explicitly  define Lai's strategy as follows.
\begin{define}\label{def:laistrat}
A code $(f,\varphi)$ uses Lai's strategy if:
\begin{itemize}
    \item for each $\mathbf{y} \in \mathcal{Y}$ there is at most one value of $k \in \mcf{K}$ such that   $\varphi(\mathcal{M}|\mathbf{y},k) > 0$,
    \item $\mathbb{I}(\mathbf{Z};K) \leq \epsilon.$
    \end{itemize}
\end{define}
With this strategy, each of Bob's possible observed sequence can only correspond to a single key.
Hence, any attempt to interlope would require Gr{\'i}ma to select an observation corresponding to the secret key shared between Alice and Bob. 
To that end, by Alice transmitting the value of the secret key in a way which obfuscates its value from Gr{\'i}ma, she ensures that Gr{\'i}ma's probability of determining which secret key is being used remains small.

For the specific code, Lai et al.~\cite{lai2009authentication} used a modified wiretap coding scheme where, in particular, they first chose an integer $n$ and distribution $\rho \in \mcf{P}_n(\mcf{X})$ such that for $X \sim \rho(x)$ $$\begin{array}{lrl}&\mathbb{I}(Y;X) - \mathbb{I}(Z;X) &> 0, \\ &|\mcf{M}||\mcf{K}| &< 2^{n \mathbb{I}(Y;X) }, \\ \text{ and } &|\mcf{K}| &< 2^{n \left[ \mathbb{I}(Y;X) - \mathbb{I}(Z;X)\right] }. \end{array}$$ 
Next, they randomly and independently selected approximately $2^{n \mathbb{I}(Y;X) }$ codewords from the type set of $\mbcf{T}_{\rho}$.
These codewords were then placed into one of $2^{n \left[ \mathbb{I}(Y;X) - \mathbb{I}(Z;X)\right] }$ bins at random, giving approximately $2^{n\mathbb{I}(Z;X)}$ codewords per bin.
Each of these bins were then associated with a particular key, and each codeword in the bin was assigned a message.
Because the capacity of the channel from Alice to Gr{\'i}ma was entirely exhausted sending the information about the message given the secret key, the secret key remained obscured from Gr{\'i}ma and yet still correlated with the message.

While one of our coding schemes will rely on Lai's strategy, as stated in Definition~\ref{def:laistrat}, we will not limit ourselves to their their coding scheme. 
Instead, we shall use a general code for the DM-BCCC which we describe in greater detail in Section~\ref{sec:back:bcc}.

\subsection{Simmons' strategy}\label{sec:back:simmons}

Simmons~\cite{Auth} considered this problem where all channels were noiseless.
Simmons' strategy, specifically, was to associate each key $k \in \set{1,\dots,2^{n\kappa}}$ with an independently and randomly chosen subset $\mbcf{X}(k) \subset \mbcf{X}$ where 
\[
\abs{\mbcf{X}(k)} = 2^{- n \kappa/2}|\mbcf{X}| = |\mcf{M}|. 
\]
For each $m \in \mcf{M}$ and $k$, Alice chooses a unique $\mbf{x} \in \mbcf{X}(k)$ to represent the message. 
Hence, the message rate is $$n^{-1} \logt |\mcf{M}| = n^{-1}\logt |\mbcf{\tilde X}| - \kappa/2 .$$ 

On the other hand, consider the scenario where Gr{\'i}ma observes $\mbf{x}$ and replaces it with $\mbf{x}'\neq \mbf{x}$. 
Having observed $\mbf{x}$, Gr{\'i}ma can narrow down the value of the key (since not all $\mbcf{X}(k)$ contain $\mbf{x}$) and use this information in the selection of $\mbf{x}'$. 
On average, there should be $\abs{\mcf{K}} ( \abs{\mbcf{X}(k)} /\abs{\mbcf{X}})^2 =  1$ value of $k$ such that $\mbcf{X}(k)$ contain both $\mbf{x}$ and $\mbf{x}'$. 
At the same time, there will be on average $\abs{\mcf{K}} ( \abs{\mbcf{X}(k)} /\abs{\mbcf{X}}) = 2^{n \kappa/2}$ values of $k$ such that $\mbcf{X}(k)$ contains $\mbf{x}$. 
Hence, on average Gr{\'i}ma should only have a $2^{-n\kappa/2}$ chance of selecting a $\mbf{x}'$ which is actually valid for the given secret key.

\subsection{Broadcast channel with confidential communications} \label{sec:back:bcc}

 Optimal codes for the DM-BCC$(p_{Y|X},p_{Z|X})$ channel were first determined by Csisz{\'a}r and K{\"o}rner~\cite{csiszar1978broadcast}, and later improved by the same authors~\cite[Chapter~17]{CK}. 
 We describe what appears in~\cite[Chapter~17]{CK}. 
 Formally, a DM-BCCC$(p_{Y|X},p_{Z|X})$ is similar to a DM-AIC$(p_{Y|X},p_{Z|X})$ where Gr{\'i}ma is forced to select Alice's transmission. 
That is, Alice's transmits sequence $\mbf{X}$, and Bob receives sequence $\mbf{Y}$, where $\mbf{Y}|\{\mbf{X} = \mbf{x}\} \sim   \mbf{p}_{Y|X}(\mbf{y}|\mbf{x})$, while Gr{\'i}ma receives sequence $\mbf{Z}$, where $\mbf{Z}|\{\mbf{X} = \mbf{x}\} \sim   \mbf{p}_{Z|X}(\mbf{z}|\mbf{x})$.

For the DM-BCCC$(p_{Y|X},p_{Z|X})$, though, Alice is attempting to send three different messages $(M_0,M_1,M_s)$, each with unique requirements. 
To wit, message $M_0$ will need to be reliably decoded by both Bob and Gr{\'i}ma, message $M_1$ reliably decoded by only Bob, and message $M_s$ reliably decoded by Bob but also kept secret from Gr{\'i}ma.  
Our direct results will show codes of this type, when used with Lai's strategy, naturally provide good authentication codes.
Because of this, the following more formal definition\footnote{Csisz{\'a}r and K{\"o}rner leave the formal definition up to the reader. These formal error definitions can be directly inferred from their code construction, which relies on~\cite[Lemma~17.14]{CK}, while the deterministic decoder requirement and leakage requirement are easily inferred as consequences of \cite[Theorem~17.13]{CK} being an extension of a simpler problem whose requirements are defined in~\cite[Definition~17.10]{CK}.} will be of use.
\begin{define}
A code $$(\tilde f, \tilde \varphi,\hat \varphi) \in (\mcf{P}(\mbcf{X}|\mcf{M}_0,\mcf{M}_1,\mcf{M}_s), \mcf{P}(\mcf{M}_0,\mcf{M}_1,\mcf{M}_s|\mbcf{Y}),\mcf{P}(\mcf{M}_0|\mbcf{Z}))$$ is a $(r_0,r_1,r_s,\epsilon,n)$ code for the DM-BCCC$(p_{Y|X},p_{Z|X})$ if the following are satisfied
\begin{itemize}
    \item $|\mcf{M}_0| \geq 2^{nr_0}$, $|\mcf{M}_1| \geq 2^{nr_1}$, $|\mcf{M}_s| \geq 2^{nr_s}$ ,
    \item $\tilde \varphi(m_0,m_1,m_s|\mbf{y}) \in \{0,1\}$ for each $(m_0,m_1,m_s,\mbf{y}) \in (\mcf{M}_0,\mcf{M}_1,\mcf{M}_s,\mbcf{Y}),$
    \item $$ \mathbb{E} \left[ \tilde \varphi(M_0,M_1,M_s|\mbf{Y}) \right] \geq 1 -\epsilon,$$
    \item $$ \mathbb{E} \left[ \hat \varphi(M_0|\mbf{Z}) \right] \geq 1 -\epsilon,$$
    \item $$\mathbb{I}(M_s;\mbf{Z}) \leq \epsilon.$$
\end{itemize}
\end{define}

\begin{define}
A triple of positive real numbers $(r_0,r_1,r_s)$ is achievable for the DM-BCCC$(p_{Y|X},p_{Z|X})$ if there exists a sequence of $(a_i,b_i,c_i,\epsilon_i,i)$-codes, for $i \in \mathbb{Z}^+$, such that 
$$\lim_{i \rightarrow \infty} |(a_i,b_i,c_i,\epsilon_i) - (r_0,r_1,r_s,0) | = 0.  $$
\end{define}

Under these definitions, Csisz{\'a}r and K{\"o}rner proved the following theorem~\cite[Theorem~17.13]{CK}.
\begin{theorem}\label{thm:ck1713} (\cite[Theorem~17.13]{CK})
The triple $(r_0,r_1,r_s)$ is achievable if and only if  
\begin{align*}
r_0 + r_s + r_1 &\leq \mathbb{I}(Y;U|W ) + \min \left( \mathbb{I}(Y;W), \mathbb{I}(Z;W) \right)\\
r_s &\leq  \mathbb{I}(Y;U|W)- \mathbb{I}(Z;U|W)  \\
r_0 &\leq \min \left( \mathbb{I}(Y,W), \mathbb{I}(Z;W)\right) ,
\end{align*}
for some RVs $U$ and $W$ such that $\abs{\mcf{U}} \leq (\abs{\mcf{X}}+1)(\abs{\mcf{X}}+3)$, $\abs{\mcf{W}} \leq \abs{\mcf{X}}+3$, and $W \markov U \markov X \markov (Y,Z)$. 
\end{theorem}

As will be shown later, a code for the DM-BCCC$(p_{Y|X},p_{Z|X})$, where $K$ is transmitted using $M_s$ and $M$ is transmitted using $M_1$ naturally provides a coding scheme which satisfies Lai's strategy (see Definition~\ref{def:laistrat}). 
By having Alice send the secret key with the secret message, Bob can decode the transmitted value and check against his own copy of the secret key to determine validity.
At the same time, Gr{\'i}ma will not be able to gain much information about the secret key since $M_s$ is designed to be secret.
Thus, Gr{\'i}ma's probability of intruding in the system should be around $2^{-n\kappa}$ since he gained no information about the key, and at the same time each of Bob's observations, which Gr{\'i}ma may choose, correspond to only a single key. 

To conclude this section, we simplify the region in Theorem~\ref{thm:ck1713} for triples of the form $(0,r_1,r_s).$ 
Such a step is prudent since, for our purposes, there is nothing to be gained by designing our coding scheme around transmission of a message that Gr{\'i}ma can decode.  
\begin{cor}\label{cor:ck1713}
The triple $(0,r_1,r_s)$ is achievable if and only if  
\begin{align*}
r_s + r_1 &\leq \mathbb{I}(Y;U,W ) \\
r_s &\leq  \mathbb{I}(Y;U|W)- \mathbb{I}(Z;U|W)  
\end{align*}
for some RVs $U$ and $W$ such that $\abs{\mcf{U}} \leq (\abs{\mcf{X}}+1)(\abs{\mcf{X}}+3)$, $\abs{\mcf{W}} \leq \abs{\mcf{X}}+3$, and $W \markov U \markov X \markov (Y,Z)$. 
\end{cor}
\begin{IEEEproof}
First note that $(0,r_1,r_s)$ is achievable if and only if $(r_1,r_s) \in \mcf{R}$, where $\mcf{R}$ is the set of $(r_1,r_s)$ for which
\begin{align*}
 r_s + r_1 &\leq \mathbb{I}(Y;U|W ) + \min \left( \mathbb{I}(Y;W), \mathbb{I}(Z;W) \right)\\
r_s &\leq  \mathbb{I}(Y;U|W)- \mathbb{I}(Z;U|W)  
\end{align*}
for some $U$ and $W$ such that $\abs{\mcf{U}} \leq (\abs{\mcf{X}}+1)(\abs{\mcf{X}}+3)$, $\abs{\mcf{W}} \leq \abs{\mcf{X}}+3$, and $W \markov U \markov X \markov (Y,Z)$, by Theorem~\ref{thm:ck1713}.
To prove the corollary, we must demonstrate that $\mcf{R} = \mcf{R}'$, where $\mcf{R}'$ is the set of $(r_1,r_s)$ such that 
\begin{align*}
 r_s + r_1 &\leq \mathbb{I}(Y;U,W ) \\
r_s &\leq  \mathbb{I}(Y;U|W)- \mathbb{I}(Z;U|W)  
\end{align*}
for some $U$ and $W$ such that $\abs{\mcf{U}} \leq (\abs{\mcf{X}}+1)(\abs{\mcf{X}}+3)$, $\abs{\mcf{W}} \leq \abs{\mcf{X}}+3$, and $W \markov U \markov X \markov (Y,Z)$.

To this end, it is clear that $\mcf{R} \subseteq \mcf{R}'$ since $\min \left( \mathbb{I}(Y;W), \mathbb{I}(Z;W) \right) \leq \mathbb{I}(Y;W).$

Now to show $\mcf{R}' \subseteq \mcf{R}$, let $(r_1,r_s) \in \mcf{R}'$ and further let $U$ and $W$ be the RVs guaranteed to exist via $(r_1,r_s)$ being a point in $\mcf{R}'.$ 
If $\mathbb{I}(Y;W) \leq \mathbb{I}(Z;W)$, then clearly $(r_1,r_s) \in \mcf{R}.$ 
On the other hand if $\mathbb{I}(Y;W) > \mathbb{I}(Z;W)$ then setting $\hat U = U$ and $\hat W = \emptyset$ provides
\begin{align*}
\mathbb{I}(Y;\hat U|\hat W) + \min \left( \mathbb{I}(Y;\hat W), \mathbb{I}(Z;\hat W) \right)  &= \mathbb{I}(Y;U) \\
&= \mathbb{I}(Y;U,W) \\
&\geq r_1 +r_s \\
  \mathbb{I}(Y;\hat U|\hat W)- \mathbb{I}(Z;\hat U|\hat W)  &= \mathbb{I}(Y;U) - \mathbb{I}(Z;U) \\
  &= \mathbb{I}(Y;U,W) - \mathbb{I}(Z;U,W)\\
  &= \mathbb{I}(Y;U|W) - \mathbb{I}(Z;U|W) + \mathbb{I}(Y;W) - \mathbb{I}(Z;W) \\
  &> \mathbb{I}(Y;U|W) - \mathbb{I}(Z;U|W) \\
  &\geq r_s.
\end{align*}
Furthermore $|\mcf{\hat U}| = |\mcf{U}| \leq (\abs{\mcf{X}}+1)(\abs{\mcf{X}}+3)$, $|\mcf{\hat W}| = 0 \leq \abs{\mcf{X}}+3$, and $\hat W \markov \hat U \markov X \markov (Y,Z),$
hence $(r_1,r_s) \in \mcf{R}.$

\end{IEEEproof}

\subsection{Information stabilizing random variables}\label{sec:inducing}

Converse proofs will rely heavily on~\cite[Corollary~17]{graves2018inducing}.
This result is hard to parse due to it's generality, so for presentation purposes, we specialize to the model at hand and have made all error terms equal.

\begin{theorem}(\cite[Corollary~17]{graves2018inducing})
For each code and pair of discrete memoryless channels $(p_{Y|X},p_{Z|X})$, there exists RV $T$ such that:
\begin{itemize}
    \item \begin{equation}\label{eq:sizeoft}
        n^{-1}\logt |\mcf{T}| \leq  \lambda_n, 
    \end{equation}  
    \item \begin{equation} (T,M,K) \markov \mbf{X} \markov (\mbf{Y},\mbf{Z}),
    \end{equation}
    \item \begin{equation}\label{eq:pofd}
\Pr \left( (\mbf{Y},\mbf{Z},M,K,T) \in \mcf{D}^+ \right) \geq 1 - 2^{-n\lambda_n},
\end{equation}
\end{itemize}
where  $\lambda_n = O(n^{-\frac{1}{2|\mcf{X}|\max(| \mcf{Y}|,|\mcf{Z}|)+2}} \logt n)$ and  $\mcf{D}^+$ is the set of $(\mbf{y},\mbf{z},m,k,t)$ such that
$$\begin{array}{ r l l}
 p(\mbf{y}|t) &= 2^{-\mathbb{H}(\mbf{Y}|T=t) + n\lambda_n}\\
 p(\mbf{z}|t) &= 2^{-\mathbb{H}(\mbf{Z}|T=t) + n\lambda_n}\\
 p(\mbf{y}|m,t) &= 2^{-\mathbb{H}(\mbf{Y}|M,T=t) + n\lambda_n}\\
 p(\mbf{z}|m,t) &= 2^{-\mathbb{H}(\mbf{Z}|M,T=t) + n\lambda_n}\\
p(\mbf{y}|k,t) &= 2^{-\mathbb{H}(\mbf{Y}|K,T=t) + n\lambda_n}\\
 p(\mbf{z}|k,t) &= 2^{-\mathbb{H}(\mbf{Z}|K,T=t) + n\lambda_n}\\
 p(\mbf{y}|m,k,t) &= 2^{-\mathbb{H}(\mbf{Y}|M,K,T=t) + n\lambda_n}\\
 p(\mbf{z}|m,k,t) &= 2^{-\mathbb{H}(\mbf{Z}|M,K,T=t) + n\lambda_n}\\
 p(m|t) &= 2^{-n r   + n\lambda_n}\\
 p(k|t) &= 2^{-n \kappa   + n\lambda_n}\\
 p(m,k|t) &= 2^{-n (r+\kappa)   + n\lambda_n}.
\end{array}
$$
\end{theorem}
\begin{remark}
Furthermore, we will write statements of the form $(\mbf{y},m,k,t) \in \mcf{D}^+$ in lieu of defining a new set, say $\mcf{\tilde D}^+$, that consists of all $(\mbf{y},m,k,t)$ such that
$$\begin{array}{ r l l}
 p(\mbf{y}|t) &= 2^{-\mathbb{H}(\mbf{Y}|T=t) + n\lambda_n}\\
p(\mbf{y}|m,t) &= 2^{-\mathbb{H}(\mbf{Y}|M,T=t) + n\lambda_n}\\
 p(\mbf{y}|k,t) &= 2^{-\mathbb{H}(\mbf{Y}|K,T=t) + n\lambda_n}\\
p(\mbf{y}|m,k,t) &= 2^{-\mathbb{H}(\mbf{Y}|M,K,T=t) + n\lambda_n}\\
 p(m|t) &= 2^{-n r   + n\lambda_n}\\
p(k|t) &= 2^{-n \kappa   + n\lambda_n}\\
 p(m,k|t) &= 2^{-n (r+\kappa)   + n\lambda_n}.
\end{array}
$$
To this end, if we were to define $\mcf{\tilde D}^+$ as above then 
$$\Pr \left( (\mbf{Y},M,K,T) \in \mcf{\tilde D}^+ \right) \geq \Pr \left( (\mbf{Y},\mbf{Z},M,K,T) \in \mcf{D}^+ \right) \geq 1 - 2^{-n\lambda_n},$$ 
since if there exists a $\mbf{z}$ such that $(\mbf{y},\mbf{z},m,k,t) \in \mcf{D}^+ $, then $(\mbf{y},m,k,t) \in \mcf{\tilde D}^+$.
Defining a new stabilizing set, such as $\mcf{\tilde D}^+$ above, would require us introduce notation for each of the $15$ possible non empty substring of $(\mbf{y},\mbf{z},m,k,t)$ that include $t$ (i.e., $(\mbf{y},\mbf{z},m,t)$, $(\mbf{y},\mbf{z},k,t)$, and so on). 
Instead, we emphasize the recursive nature of this result here, and opt for using a single $\mcf{D}^+$, considering it the stabilized set.

\end{remark}

These properties will be useful in constructing information theoretic necessary conditions from authentication, acting as a general Fano's inequality.
That is, where Fano's inequality uses the probability of error to derive a bound on conditional entropy, these properties will allow us to establish requirements on associated information terms using the systems requirements for authentication.
To give an example of how these properties may be used, we conclude this section by demonstrating how to derive  
\begin{equation}\label{eq:fanosalt} 
r \leq  \max_{t} \frac{1}{n}  \mathbb{I}(\mbf{Y}; M|T=t) + \tilde \lambda_n ,
\end{equation}
where $\tilde \lambda_n = 3\lambda_n - \frac{1}{n}\logt ( 1 - \epsilon - 2^{-n\lambda_n} ) ,$
for any code that satisfies 
$$1-\epsilon \leq \sum_{\mbf{y},m,k} p(\mbf{y},m,k) \varphi(m|\mbf{y},k).$$
From Equation~\eqref{eq:fanosalt} it can be further shown that 
\begin{equation}\label{eq:fanosalt2} 
r \leq  \max_{p_{X}}  \mathbb{I}(Y; X) + \tilde \lambda_n
\end{equation}
by using the chain rule for mutual information, and that conditional entropy is always strictly less than the unconditional entropy, and then making use of since $(T,M) \markov \mbf{X} \markov (\mbf{Y}, \mbf{Z}).$
Equation~\eqref{eq:fanosalt2}, where the error term is replaced with one linearly dependent upon probability of error $\epsilon$, can easily be derived starting from Fano's inequality (see, for example,~\cite[Section~7.9]{CT}); the derivation here is intended simply to provide an example of how the information stabilizing random variable can produce information theoretic necessary requirements.

Using the properties of the information stabilizing random variable $T$, Equation~\eqref{eq:fanosalt} may be derived directly from the average probability of error requirement as so;
\begin{align}
1-\epsilon &\leq \sum_{\mbf{y},m,k} p(\mbf{y},m,k) \varphi(m|\mbf{y},k) \notag \\
&= \sum_{\mbf{y},m,k,t} p(\mbf{y},m,k,t) \varphi(m|\mbf{y},k) \label{eq:ispex1} \\
&\leq 2^{-n\lambda_n} + \sum_{(\mbf{y},m,k,t) \in \mcf{D}^+} p(\mbf{y},m,k,t) \varphi(m|\mbf{y},k)  \label{eq:ispex2} \\
&= 2^{-n\lambda_n} + \sum_{(\mbf{y},m,k,t) \in \mcf{D}^+}  \frac{p(\mbf{y}|m,t)}{p(\mbf{y}|t)}  p(m|t) p(k|\mbf{y},m,t)  p(\mbf{y},t)  \varphi(m|\mbf{y},k)  \label{eq:ispex3} \\
&\leq 2^{-n\lambda_n} + \sum_{(\mbf{y},m,k,t) \in \mcf{D}^+}  2^{-\left[ n r - \mathbb{I}(\mbf{Y};M|T=t) - 3n\lambda_n \right] }  p(k|\mbf{y},m,t)  p(\mbf{y},t)  \varphi(m|\mbf{y},k)  \label{eq:ispex4} \\
&\leq 2^{-n\lambda_n} + \sum_{t}  2^{-\left[ n r - \mathbb{I}(\mbf{Y};M|T=t) - 3n\lambda_n\right]}  p(t) \label{eq:ispex5} \\
&\leq 2^{-n\lambda_n} + \max_{t}  2^{-\left[n r - \mathbb{I}(\mbf{Y};M|T=t) - 3n\lambda_n\right]}  \label{eq:ispex6} ;
\end{align}
where~\eqref{eq:ispex1} is the law of total probability;~\eqref{eq:ispex2} is because of Equation~\eqref{eq:pofd} and because $\varphi$ is a probability distribution;~\eqref{eq:ispex3} is Bayes' Theorem;~\eqref{eq:ispex4} is because 
$$ \frac{p(\mbf{y}|m,t)}{p(\mbf{y}|t)}  p(m|t)  \leq 2^{-nr + \mathbb{I}(\mbf{Y};M|T=t) + 3 n \lambda_n}$$
for all $(\mbf{y},m,k,t) \in \mcf{D}^+;$~\eqref{eq:ispex5} is by recognizing that $ 2^{-nr + \mathbb{I}(\mbf{Y};M|T=t) + 3n \lambda_n}$ is not dependent upon $m,k$ or $\mbf{y};$ and~\eqref{eq:ispex6} is because the maximum is greater than the average.  
\section{Main Contributions} \label{sec:main_cont}

Our major contribution, Theorem~\ref{thm:WAAR}, is the characterization of the typical authentication capacity region. 
Before presenting this result, it will be helpful to first present a number of intermediary results and explain their relevance. 
This is true for both the direct result (that a particular set of $(r,\alpha,\kappa)$ can be achieved) and the converse result (that only these $(r,\alpha,\kappa)$ can be achieved).

We start with the direct results where we shall show that all $(r,\alpha,\kappa) \in \mcf{C}_{\mathrm{TA}}(p_{Y|X},p_{Z|X})$ can be achieved using the composition of two codes.
In particular, these two codes are distinct in the method by which they exploit the secret key to generate authentication rate. 
For the first code, the authentication rate will be derived from the ability of Alice to send secure information over the channel.
The second code, on the other hand, will derive the authentication rate by exploiting Alice and Bob's shared resource (the secret key) in such a way that is insensitive to the channel. 
Thus, these two direct codes can, in some abstract sense, be thought of as a code to exploit the security of the channel and a code to exploit the security of the source.

For the code that exploits the security of the channel, we opt to use a DM-BCCC$(p_{Y|X},p_{Z|X})$ code in conjunction with Lai's strategy.
Doing so leads to the following inner bound on the typical authentication capacity region, $\mcf{C}_{\mathrm{TA}}(p_{Y|X},p_{Z|X})$. 
\begin{theorem} \label{thm:lai}
For positive real numbers $r,\alpha,\kappa$ if 
\begin{align*}
 r+ \alpha &\leq \mathbb{I}(Y;U,W)   \\
\alpha &\leq \mathbb{I}(Y;U|W) - \mathbb{I}(Z,U|W) \\
\alpha - \kappa &\leq  0 ,
\end{align*} 
for some RVs $X,U,W$ such that $(U,W) \markov X \markov (Y,Z),$
then $(r,\alpha,\kappa) \in \mcf{C}_{\mathrm{TA}}(p_{Y|X},p_{Z|X})$.
\end{theorem}

\noindent See Appendix~\ref{sec:proof_lai} for proof.

The rate region derived in Theorem~\ref{thm:lai} is actually what one should intuitively expect.
That is, a DM-BCCC code allows for the transmission of a secure $n(\mathbb{I}(Y;U|W) - \mathbb{I}(Z,U|W))$-bit message.
By transmitting the secret key with this secure message, it ensures that Gr{\'i}ma does not learn any information about the key, and hence his observation does not improve his probability of guessing the key.
At the same time, Gr{\'i}ma must accurately guess the key, since Lai's strategy sees that each of Gr{\'i}ma's possible choices for Bob's observation correspond to only a single secret key.  
Thus, Gr{\'i}ma, with no further knowledge of the secret key, would have around a $2^{-n\kappa}$ chance of guessing the key, as long as $\kappa < \mathbb{I}(Y;U|W) - \mathbb{I}(Z,U|W)$.
Of course, for $\kappa \geq \mathbb{I}(Y;U|W) - \mathbb{I}(Z,U|W) $ it is always possible to only use a subset of the secret key bits at a given time, hence the restriction should be considered on the authentication rate instead of the key consumption rate. 

Of interest to note, here, is that using the DM-BCCC code with Lai's strategy requires there to be a trade-off between the authentication rate and the message rate, since more bits being used for the secret message means less bits for the non-secret message. 
This is reflected in the upper bound on the sum of message rate and authentication rate.

For the second coding scheme, we will exploit the shared resource, i.e.~the secret key, independently of the channel.
In particular, we modify Simmons'~\cite{Auth} coding scheme into a universally composable code. 
\begin{theorem}\label{lem:-t}
If $(r,\alpha, \kappa) \in \mcf{C}_{\mathrm{TA}}(p_{Y|X},p_{Z|X})$ then $$(r-\beta,\alpha+\beta, \kappa+ 2\beta) \in \mcf{C}_{\mathrm{TA}}(p_{Y|X},p_{Z|X})$$ for all non-negative $\beta < r$.
\end{theorem}

\noindent See Appendix~\ref{app:cc2} for proof.

Practically, one may think of the coding scheme which accomplishes the above rate region as follows.
Given a starting code, take $n\beta$-bits assigned to transmitting the message and reallocate them for authentication. 
Next, independently for each $n(r-\beta)$-bit message, randomly choose an isomorphic $n\beta$ to $n\beta$-bit mapping.  
Now, for communication, apply the appropriate isomorphic mapping to the first $n\beta$ bits of additional secret key and add to the result (bitwise modulo $2$) the remaining $n\beta$ bits of additional secret key.
The $n\beta$-bit sequence that results from the addition is then sent using the $n\beta$ reallocated for the purpose of authentication.
This addition acts as a one-time pad ensuring the output of the isomorphic mapping is secret from Gr{\'i}ma.
At the same time, if Gr{\'i}ma were to change the message, he would also need to choose the unique $n\beta$-bit sequence relating to his chosen message and the additional $2n\beta$ bits of secret key.
Gr{\'i}ma having no information about what that $n\beta$-sequence should be, then, should have at most a $2^{-n\beta}$ probability of guessing the correct sequence.

Clearly, this type of code sacrifices message rate and key consumption rate in order to increase the authentication rate. 
In particular, an increase of $\beta$ in the authentication rate requires an increase of $2\beta$ in the key consumption rate. 
Intuitively, two bits of secret key are consumed for every one bit of authentication needed.  
This differs from Theorem~\ref{thm:lai} where $\alpha = \kappa$ up to a given threshold.
On the other hand, similar to Theorem~\ref{thm:lai}, the authentication rate and the message rate satisfy a linear relationship.
That is, the sum of the message rate and authentication rate is preserved. 

These results are sufficient to establish the direct portion of $\mcf{C}_{\mathrm{TA}}(p_{Y|X},p_{Z|X}),$ thus we move on to results which support the converse. 
For the converse, the following intermediary result is needed. 

\begin{lemma}\label{lem:c:2}
There exists a function $\zeta : \mathbb{R}^+ \times \mathbb{Z}^+ \rightarrow \mathbb{Z}^+$, where $\lim_{\subalign{a &\rightarrow 0^+\\ n &\rightarrow \infty}} \zeta(a,n) = 0,$ such that 
\begin{align}
 \alpha &\leq   n^{-1}\min \left( ~\mathbb{I}(\mbf{Y};K)~, ~\mathbb{H}(K|\mbf{Z})~ \right) + \zeta(\epsilon,n) \label{eq:lem:c:2:stat}
\end{align}
for all $(r,\alpha,\kappa, \epsilon, n)$-TA codes for DM-AIC$(p_{Y|X},p_{Z|X}).$
\end{lemma}

\noindent See Appendix~\ref{app:c} for proofs.

These bounds are due to the traditional ``impostor'' and ``substitution'' attacks.
In particular, the upper bound of $\mathbb{I}(\mbf{Y};K)$ is a result of the authentication system needing to be robust against attacks where the adversary chooses Bob's observation $\mbf{y}$ according to  $p_{\mbf{Y}}(\mbf{y})$.
On the other hand, the upper bound of $\mathbb{H}(K|\mbf{Z})$ is a result of having to defend against attacks where the adversaries chooses $\mbf{y}$ as a function of their observation $\mbf{z}$ according to $\sum_{k}p(\mbf{y}|k) p(k|\mbf{z}).$

Similar bounds exist in literature for average authentication codes.
In particular,
\begin{align}
 -\logt \omega_{f,\varphi} \leq  \min \left( ~\mathbb{I}(\mbf{Y};K)~, ~\mathbb{H}(K|\mbf{Z})~ \right) \label{eq:results:bad} 
\end{align}
for any code $(f,\varphi)$, where the $\mathbb{I}(\mbf{Y};K)$ upper bound is from Simmons'~\cite[Theorem~3]{Auth}, and the $\mathbb{H}(K|\mbf{Z})$ upper bound is a consequence of having to defend against attacks in which Gr{\'i}ma chooses the most likely key given his observation and attacks under the assumption it is the correct key and hence,
\begin{align}
\omega_{f,\varphi} &\geq \sum_{\mbf{z}} p(\mbf{z}) \max_{k} p(k|\mbf{z}) \\
&= \sum_{\mbf{z}} p(\mbf{z}) 2^{\max_{k} \logt p(k|\mbf{z})} \\
&\geq \sum_{\mbf{z}} p(\mbf{z}) 2^{\sum_{k} p(k|\mbf{z}) \logt p(k|\mbf{z})} \\
&\geq  2^{\sum_{k,\mbf{z}} p(k,\mbf{z}) \logt p(k|\mbf{z})} = 2^{-\mathbb{H}(K|\mbf{Z})}.
\end{align}
Lemma~\ref{lem:c:2}, therefore, extends these previous conclusions to the strictly not smaller set of typically achievable $(r,\alpha,\kappa)$. 
This is somewhat unfortunate, though, as Lemma~\ref{lem:c:2} will provide asymptotically tight bounds (as shown by the upcoming Theorem~\ref{thm:WAAR}).
Hence, if the average authentication region is in general a strict subset of the typical authentication region (which we conjecture), then it follows that~\eqref{eq:results:bad} is loose.

These preceding results, plus classic well known techniques (essentially Fourier-Motzkin elimination, Csisz{\'a}r sum identity, and Fenchel-Eggleston-Carath{\'e}odory theorem), combine to prove the main theorem.
\begin{theorem}\label{thm:WAAR}
The typical authentication capacity region, $\mcf{C}_{\mathrm{TA}}(p_{Y|X},p_{Z|X})$, is the set of $(r,\alpha, \kappa)$ such that 
\begin{align*}
r + \alpha &\leq \mathbb{I}(Y;U,W)  \\
2 \alpha - \kappa  &\leq   \mathbb{I}(Y;U|W) -  \mathbb{I}(Z;U|W)   \\
\alpha - \kappa &\leq 0 
\end{align*}
for some random variables $X,U,W$ such that $W \markov U \markov X \markov (Y,Z)$ and $\abs{\mcf{U}} = (\abs{\mcf{X}}+2)(\abs{\mcf{X}}+1)$ and $\abs{\mcf{W}} = \abs{\mcf{X}}+2$.
\end{theorem}
\noindent See Appendix~\ref{app:WAAR_proof} for proof.

Fixing a $W,U,X$, the resulting region is best viewed in terms of the cost of authentication.
Specifically, there exists a threshold\footnote{It is perhaps best to view the second inequality as $\alpha + (\alpha - \kappa) \leq \mathbb{I}(Y;U|W) -  \mathbb{I}(Z;U|W)$.} $( \mathbb{I}(Y;U|W) -  \mathbb{I}(Z;U|W))$ below which every bit of authentication costs one bit of message rate and one bit of key consumption rate. 
Above this threshold, every bit of authentication costs one bit of message rate and two bits of key consumption rate until no message rate remains. 
It is not surprising that authentication rate requires key consumption rate, more interesting is that authentication and message rate are actually a shared resource.

Of course, by allowing $W,U,X$ to vary means that these trade-offs may not be necessarily true depending on the exact message rate, authentication rate, and key consumption rate.
An important threshold in this regards is the threshold for bits of authentication below which it is possible to achieve the max sum of message and authentication rate and still have the authentication rate equal to the key consumption rate.
Specifically, this value is
$$\max_{W: W \markov X \markov (Y,Z)} \mathbb{I}(Y;X|W) - \mathbb{I}(Z;X|W),$$
where the distribution of $X$ maximizes $\mathbb{I}(Y;X).$
Alternatively, in environments where maximizing the authentication and message rates is critical, this threshold represents the point at which further bits of authentication cost twice as much in key consumption.
Regardless, this threshold represents where authentication rate costs the least to achieve.

Finally, it is important to note that the fact that authentication rate and message rate share a finite resource must also hold true under the average authentication metric. 
Indeed, obviously if $(r,\alpha,\kappa) \in \mcf{C}_{\mathrm{AA}}(p_{Y|X},p_{Z|X})$, then $(r,\alpha,\kappa) \in \mcf{C}_{\mathrm{TA}}(p_{Y|X},p_{Z|X})$.
Furthermore, as Part I demonstrated, the sum of message rate and authentication rate can equal the channel's capacity under the average authentication metric.
Therefore, it must follow that, even under the average authentication measure, increasing the authentication rate past a given threshold must also decrease the message rate.

\section{Examples} \label{sec:examples}
We now provide some numerical examples in order to illustrate the trade-offs between the three parameters that make up the typical authentication capacity region.
The case where both $p_{Y|X}$ and $p_{Z|X}$ are \emph{binary symmetric channels} (BSC) is considered.
That is, if $p_{Y|X}$ is a BSC with parameter $\lambda_t \in [0,1/2]$, then $\mcf{X} = \mcf{Y} = \set{0,1}$ and $p_{Y|X}(0|1) = p_{Y|X}(1|0) =  \lambda_t$. 
We shall use $\lambda_t$ throughout this section to represent the BSC parameter of $p_{Y|X}$ and $\lambda_q$ to represent the BSC parameter of $p_{Z|X}$.

Since both channels are BSC, the channel with the larger $\lambda$ value will be stochastically degraded with respect to the channel with the smaller $\lambda$. 
In the case that $\lambda_t \leq \lambda_q$, then  $\mathbb{I}(Y;W) \geq \mathbb{I}(Z;W)$ and $\mathbb{I}(Y;X|W) \geq \mathbb{I}(Z;X|W)$ for all $W \markov X \markov (Y,Z)$ by the data processing inequality, hence the region dictated by Theorem~\ref{thm:WAAR} can be simplified to 
\begin{align*}
r + \alpha &\leq \mathbb{I}(Y;X)  \\
2 \alpha - \kappa  &\leq   \mathbb{I}(Y;X) -  \mathbb{I}(Z;X)   \\
\alpha - \kappa &\leq 0 .
\end{align*}
On the other hand, when $\lambda_q \leq \lambda_t$, this same property simplifies the region to
\begin{align*}
r + \alpha &\leq \mathbb{I}(Y;X)  \\
2 \alpha - \kappa  &\leq   0  \\
\alpha - \kappa &\leq 0 .
\end{align*}

Three different plots are annotated to illustrate various considerations for the region, including the trade-off between message rate and typical authentication, the efficiency of consumed key material, and the effects of main channel quality including both the more noisy and less noisy regimes.

\begin{figure}
	\centering
	\includegraphics[scale=.5]{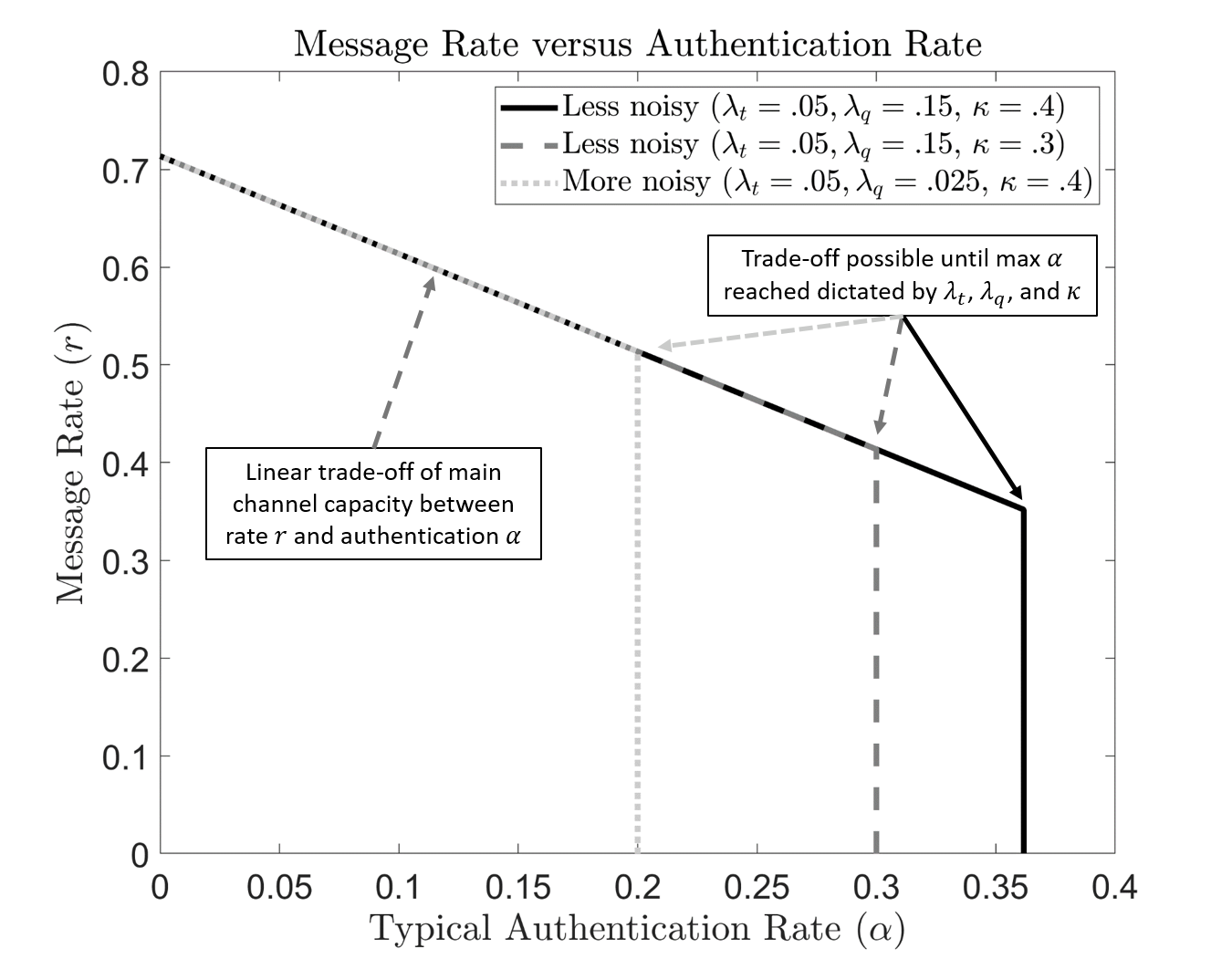}
	\caption{Message rate ($r$) vs. authentication rate ($\alpha$) for both less noisy and more noisy main channel for the typical authentication capacity region (Theorem \ref{thm:WAAR}).}
	\label{fig:RvsAlpha}
\end{figure}

A significant result present in both capacity regions of Theorem \ref{thm:lai} and Theorem \ref{thm:WAAR} is that communication and authentication must share the main channel capacity.
Fig. \ref{fig:RvsAlpha} depicts the trade-off for both the more noisy and less noisy channel cases.
The linear trade-off between $r$ and $\alpha$ is ultimately limited by the channel or key rate.
In the first less noisy case, $\alpha$ is limited by the secrecy capacity of the channel pair, while in the second case, it is limited by the amount of key available since $\kappa=.3$ is less than the secrecy capacity of the channel.
Finally when the main channel is more noisy, $\alpha$ is limited by half the key $\kappa/2$.
The ability to achieve a nonzero authentication rate in such cases is due to the incorporation of Simmons' noiseless strategy in our code, unlike Lai's region (Theorem \ref{thm:lai}) where no authentication is possible.
The region is clearly improved when secrecy capacity is available and there is enough key material to take full advantage of it.

\begin{figure}
	\centering
	\includegraphics[scale=.5]{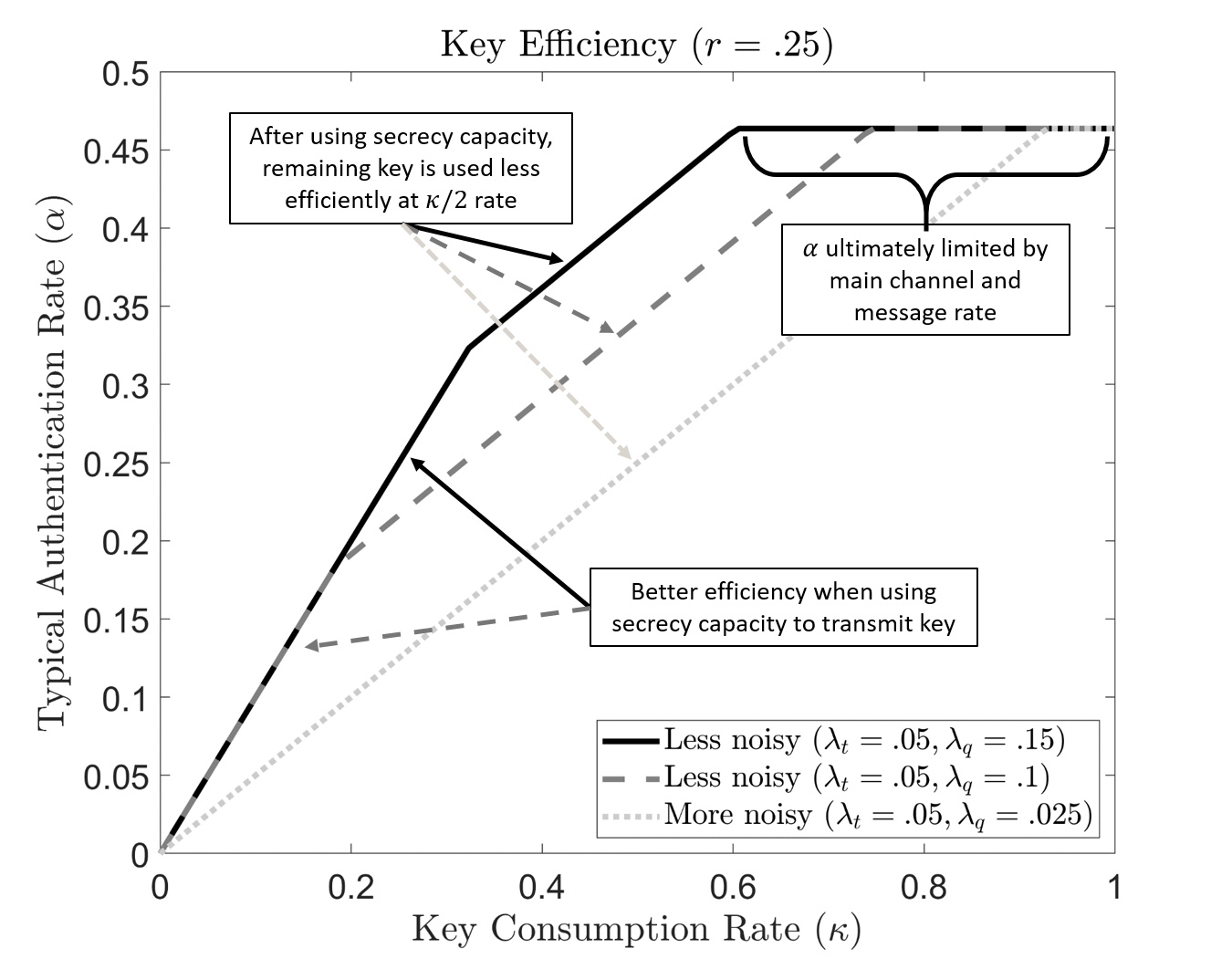}
	\caption{The amount of authentication rate gained per increase in key consumption rate is better when secrecy capacity is nonzero. Curves obtained for the typical authentication capacity region (Theorem \ref{thm:WAAR}).}
	\label{fig:keyEfficiency}
\end{figure}

With that in mind, we next explore how efficient key use is in terms of the amount of authentication rate achieved by each additional bit of key for a few different scenarios.
Key consumption is most efficient when secrecy capacity is available and used fully.
In other words, as much key as possible should be sent using the secrecy provided by the channel rather than by Simmons' strategy.
In fact, using secrecy capacity is twice as efficient, as depicted in Fig. \ref{fig:keyEfficiency}.
The two less noisy cases show that each bit of key increases the authentication rate the same amount.
However, once the secrecy capacity has been depleted, the effect of each additional bit of key is halved since the less efficient Simmons' scheme must be used instead.
For a more noisy main channel, no secrecy capacity is available, so only Simmons' strategy is used, maintaining a constant efficiency of $1/2$ for all key consumption rates.
Ultimately, though, authentication in all cases is limited by the main channel capacity and desired message rate.

\begin{figure}
	\centering
	\includegraphics[scale=.5]{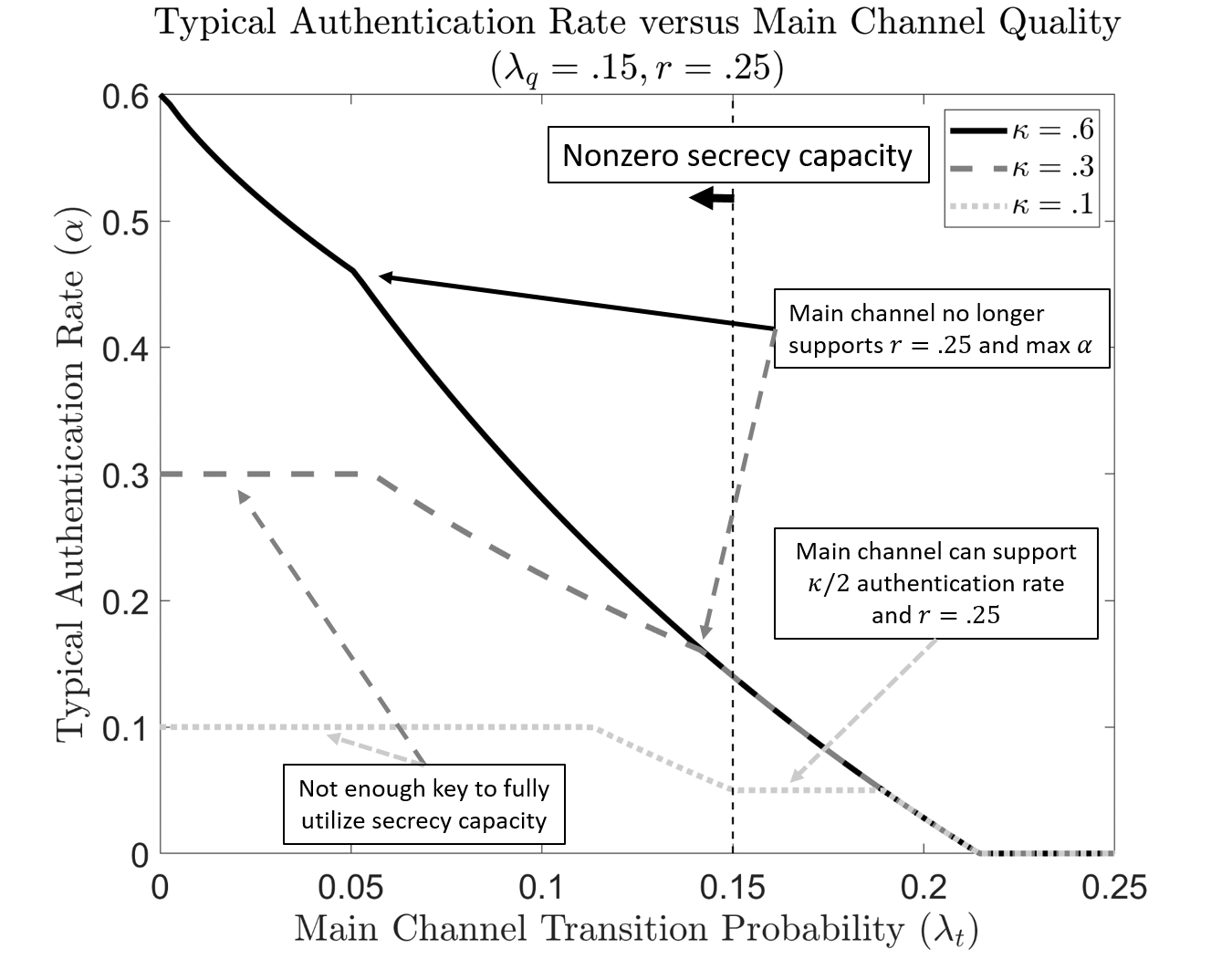}
	\caption{Authentication capabilities decrease with worsening channel conditions. Curves obtained for the typical authentication capacity region (Theorem \ref{thm:WAAR}).}
	\label{fig:varyMainChannel}
\end{figure}

Next, the effect of main channel quality on the amount of authentication possible for different key consumption rates is shown in Fig. \ref{fig:varyMainChannel}.
As one would expect, lower authentication rates are achievable for decreasing main channel quality.
As seen especially in the $\kappa=.3$ and $\kappa=.1$ cases, authentication rate is always limited by the amount of key material possessed even when the amount of secrecy capacity available exceeds it.
In the plot, this is demonstrated by the flat portion of the curves that show that the channel pair can accommodate the entire key, producing $\alpha=\kappa$, until the main channel worsens and secrecy drops below $\kappa$ which then becomes the limiting factor.
As the nonzero secrecy capacity point is approached, even though secrecy capacity is available, the message rate takes up a large portion of the main channel capacity, limiting the authentication rate further.
Once secrecy capacity is lost, only the $\kappa=.1$ case achieves the maximum rate of $\alpha=\kappa/2$.
The three cases eventually converge as a result of the decreasing main channel quality and the inability to sustain the desired message rate and authentication rate until both cannot be supported.

\appendices

\section{Proof of Theorem~\ref{thm:lai}}\label{sec:proof_lai}

We will use DM-BCCC codes (Definition~\ref{def:dm-bccc}) with Lai's Strategy (Definition~\ref{def:laistrat}) to prove Theorem~\ref{thm:lai}.
Specifically, we will first highlight two key benefits of Lai's Strategy. 
Next, in Appendix~\ref{app:trans_code}, we transform codes for the DM-BCCC$(p_{Y|X},p_{Z|X})$ into codes using Lai's Strategy for the DM-AIC$(p_{Y|X},p_{Z|X})$, and then use the properties of Lai's Strategy and the original DM-BCCC codes to derive the relevant operational measures. 
Finally, in Appendix~\ref{app:proof_lai_2}, we use these transformed codes in conjunction with Corollary~\ref{cor:ck1713} to prove Theorem~\ref{thm:lai}.

\subsection{Key features of Lai's Strategy} \label{app:lskf}

The first major benefit of using Lai's strategy is that it simplifies $\omega_{f,\varphi}.$ 
Indeed, by ensuring that for each $\mbf{y}$ there is at most a single $k$ such that $\varphi(\mbf{!}|\mbf{y},k) \neq 1$ it forces Gr{\'i}ma to know the exact value of $k$ to interlope. 
\begin{lemma}\label{lem:lai_strat1}
If $(f,\varphi)$ satisfies Lai's strategy then
$$ \omega_{f,\varphi}(\mbf{z},m,k) \leq \tilde \psi(k|\mbf{z}) ,$$
where 
$$
\tilde \psi(a |\mbf{z}) = \begin{cases}
\sum_{\mbf{y} : \varphi(\mcf{M}|\mbf{y},k) >0 } \psi(\mbf{y}|\mbf{z}) &\text{if } a\in \mcf{K} \\
\sum_{\mbf{y} : \varphi(\mbf{!}|\mbf{y},k) >0 } \psi(\mbf{y}|\mbf{z}) &\text{otherwise} 
\end{cases}$$ and $\tilde \psi \in \mcf{P}(\{\mcf{K},\mbf{!}\}|\mbcf{Z}).$
\end{lemma}
\begin{IEEEproof}
This first part of the lemma follows near immediately from definitions, 
\begin{align}
    \omega_{f,\varphi} (\mathbf{z},m,k) &= \sum_{\mbf{y}} \psi(\mathbf{y}|\mathbf{z}) \varphi(\mathcal{M}-\{m\}|\mathbf{y},k) \nonumber \\
    &\leq \sum_{\mbf{y}} \psi(\mathbf{y}|\mathbf{z}) \varphi(\mathcal{M}|\mathbf{y},k)  \\
    &= \tilde \psi(k|\mathbf{z}), \label{eq:psi}
\end{align}
where the last line follows because $\varphi$ is deterministic from Lai's Strategy. 

Next, to prove that $\tilde \psi$ is a valid probability distribution, we must show that $\sum_{k \in \{\mcf{K},\mbf{!}\}} \tilde \psi(k|\mbf{z}) = 1$ for all $\mbf{z}$.
Consider 

\begin{align}
\sum_{k \in \{ \mcf{K},\mbf{!}\}} \tilde \psi (k | \mbf{z}) &= \sum_{\mbf{y}} \psi(\mbf{y}|\mbf{z}) \idc{ \varphi(\mbf{!}|\mbf{y},k)>0} + \sum_{k \in \mcf{K}} \sum_{\mbf{y}} \psi(\mbf{y}|\mbf{z}) \idc{ \varphi(\mcf{M}|\mbf{y},k)>0}\\
&= \sum_{\mbf{y}} \psi(\mbf{y}|\mbf{z}) \left(\idc{\varphi(\mbf{!}|\mbf{y},k)>0}+\sum_{k \in \mcf{K}}   \idc{ \varphi(\mcf{M}|\mbf{y},k)>0}\right)\label{eq:lai_strat1:ls}\\
&= \sum_{\mbf{y}} \psi(\mbf{y}|\mbf{z})  = 1,
\end{align}
where Equation~\eqref{eq:lai_strat1:ls} is because Lai's Strategy requires that there is one, and only one, value of $a \in \{\mcf{M},\mbf{!}\}$ such that $\varphi(a|\mbf{y},k) = 1$ for each $\mbf{y}$ and $k$, proving the assertion.
\end{IEEEproof}

Another benefit of Lai's Strategy is that the information about the key is sent privately. 
In fact, keeping the mutual information allows for use of the following lemma.
\begin{lemma}\label{lem:lai_strat2}
If $\mathbb{I}(A;B) \leq c$ then 
$$\Pr\left( p(A|B) > p(A) 2^{n c} \right) \leq \frac{1}{n} + \frac{1}{nc}.$$
\end{lemma}
\begin{IEEEproof}
Let $\mcf{Q}$ be the set of $(a,b)$ such that $p(a|b)>p(a)2^{nc}$, furthermore let $\mcf{\hat S}$ be the subset of $(a,b)$ such that $(a,b) \notin \mcf{Q}$ and $p(a|b) \geq p(a)$, and $\mcf{\tilde S}$ be the subset of $(a,b)\notin \mcf{Q} $ such that $p(a|b) < p(a).$
Note that $\mcf{Q} \cup \mcf{\hat S} \cup \mcf{\tilde S} = \mcf{A} \times \mcf{B}.$

Now, if  $\mathbb{I}(A;B) \leq c$, then 
\begin{align}
c &= \sum_{\mcf{Q}} p(a,b) \logt \frac{p(a|b)}{p(a)} + \sum_{\mcf{\hat S}} p(a,b) \logt \frac{p(a|b)}{p(a)} + \sum_{\mcf{\tilde S}} p(a,b) \logt \frac{p(a|b)}{p(a)} \label{eq:lai_strat2:c=}
\end{align}
follows by expanding the definition of mutual information, and organizing the summation terms into the different sets. 
The summation over each set can be lower bounded in a unique way:  $\logt \frac{p(a|b)}{p(a)} \geq n c$ for all terms in $\mcf{Q}$; and
$$ p(a,b) \logt \frac{p(a|b)}{p(a)} \geq - p(a,b) \frac{p(a)}{p(a|b)}  = p(a)p(b) $$
for all terms in $\mcf{\hat S}$; finally $p(a,b) \logt \frac{p(a|b)}{p(a)}$ is negative for all terms in $\mcf{\tilde S}.$
Using these observations, then
\begin{align}
c  &\geq \Pr \left( (A,B) \in \mcf{Q} \right) nc  - 1  \label{eq:lai_strat2:fin}
\end{align}
follows.
Solving Equation~\eqref{eq:lai_strat2:fin} for $\Pr \left( (A,B) \in \mcf{Q} \right) $ proves the lemma. 
\end{IEEEproof}

\subsection{Transforming DM-BCCC codes into DM-AIC codes} \label{app:trans_code}

\begin{theorem}\label{thm:pamtransform}
If $$(\tilde f, \tilde \varphi,\hat \varphi) \in \mcf{P}(\mbcf{X}|\mcf{M}_0,\mcf{M}_1,\mcf{M}_s) \times \mcf{P}(\mcf{M}_0,\mcf{M}_1,\mcf{M}_s|\mbcf{Y})\times \mcf{P}(\mcf{M}_0|\mbcf{Z})$$
is a code $(r_0,r_1,r_s,\epsilon,n)$-code for the DM-BCCC$(p_{Y|X},p_{Z|X})$, then 
$$(f,  \varphi) \in \mcf{P}(\mbcf{X}|\mcf{M},\mcf{K}) \times  \mcf{P}(\{\mcf{M},\mbf{!}\}|\mbcf{Y},\mcf{K}),$$
where $\mcf{K} = \mcf{M}_s$, $\mcf{M} = \mcf{M}_1$, and
\begin{align}
f( \mbf{x} | m,k) &= \frac{1}{|\mcf{M}_0|} \sum_{m_0 \in \mcf{M}_0} \tilde f(\mbf{x}|m_0,m,k) \\
\varphi( a | \mbf{y},k) &= \begin{cases} \frac{1}{|\mcf{M}_0|} \sum_{m_0 \in \mcf{M}_0} \tilde \varphi(m_0,a,k|\mbf{y}) &\text{if } a \neq \mbf{!}\\
1-\sum_{\substack{m_0 \in \mcf{M}_0\\ m_1 \in \mcf{M}_1}} \tilde \varphi(m_0,m_1,k|\mbf{y}) &\text{otherwise},
\end{cases}
\end{align}
is a $(r_1,r_s-2 \epsilon,r_s,\epsilon + \frac{1}{n} + \frac{1}{n\epsilon} + 2^{-n\epsilon},n)$-TA code for the DM-AIC$(p_{Y|X},p_{Z|X})$. 
\end{theorem}
\begin{IEEEproof}
First, the message rate is $r_1$, the key consumption rate is $r_s$, and the blocklength is $n$, and the probability of message error is $\epsilon$ since $\mcf{M} = \mcf{M}_1$, $\mcf{K} = \mcf{M}_s$, and the block-length has not changed, and the code has not been changed, respectively. 
This leaves in question the authentication rate and the authentication fault tolerance.

For the authentication rate and authentication fault tolerance, let $\mcf{A}$ represent the set of all $\mbf{z},k$ such that $p(k|\mbf{z}) > p(k) 2^{n\epsilon}.$
Now observe that
\begin{align}
&\max_{\psi\in\mathcal{P}(\mbcf{Y}|\mbcf{Z})} \Pr \left(- n^{-1}\logt \omega_{f,\varphi} (\mathbf{Z},M,K) < r_s - 2\epsilon \right) \notag \\
&\quad \leq \max_{\psi\in\mathcal{P}(\mbcf{Y}|\mbcf{Z})} \Pr \left(- n^{-1}\logt \psi( K|\mbf{Z})  < r_s - 2\epsilon \right) \label{eq:proof_lai:type_auth_2sums:-1} \\
&\quad = \max_{\psi\in\mathcal{P}(\mbcf{Y}|\mbcf{Z})} \sum_{(k,\mbf{z}) \in \mcf{A}} p(\mbf{z},k) \idc{- n^{-1}\logt \psi(k|\mbf{z})  < r_s - 2\epsilon}  \notag \\
&\quad \quad + \sum_{( k,\mbf{z}) \notin \mcf{A}} p(\mbf{z},k) \idc{- n^{-1}\logt \psi(k|\mbf{z})  < r_s - 2\epsilon} , \label{eq:proof_lai:type_auth_2sums}
\end{align}
where Equation~\eqref{eq:proof_lai:type_auth_2sums:-1} follows directly from Lemma~\ref{lem:lai_strat1}.
The sums in Equation~\eqref{eq:proof_lai:type_auth_2sums} can be handled separately with some ease.
First, 
\begin{align}
&\sum_{(k,\mbf{z}) \in \mcf{A}} p(\mbf{z},k) \idc{- n^{-1}\logt \psi(k|\mbf{z})  < r_s - 2\epsilon} \notag\\
&\quad \leq \Pr \left( ( K, \mbf{Z}) \in \mcf{A} \right) \leq \frac{1}{n} + \frac{1}{n\epsilon} \label{eq:proof_lai:sum1}
\end{align}
follows directly from Lemma~\ref{lem:lai_strat2}.
On the other hand,
\begin{align}
&\sum_{( k,\mbf{z}) \notin \mcf{A}} p(\mbf{z},k) \idc{- n^{-1}\logt \psi(k|\mbf{z})  < r_s - 2\epsilon}\notag \\
&\quad \leq  \sum_{( k,\mbf{z}) \notin \mcf{A}} 2^{-nr_s + n\epsilon } p(\mbf{z}) 2^{nr_s - 2n\epsilon} \psi(k | \mbf{z}) \label{eq:proof_lai:sum2:-1} \\
&\quad \leq 2^{-n\epsilon} \label{eq:proof_lai:sum2}
\end{align}
where~\eqref{eq:proof_lai:sum2:-1} follows because $p(k|\mbf{z}) \leq p(k) 2^{n\epsilon} \leq 2^{-nr_s + n\epsilon}$ for each $(k,\mbf{z}) \notin \mcf{A}$ and because $$\idc{- n^{-1}\logt \psi(k|\mbf{z})  < r_s - 2\epsilon} \leq 2^{nr_s - 2n\epsilon} \psi(k | \mbf{z}) $$  since $\psi$ only produces positive values; and~\eqref{eq:proof_lai:sum2} follows because $\psi$ is a probability distribution. 
Combining Equations~\eqref{eq:proof_lai:sum1} and~\eqref{eq:proof_lai:sum2} yields 
\begin{equation}
\max_{\psi\in\mathcal{P}(\mbcf{Y}|\mbcf{Z})} \Pr \left(- n^{-1}\logt \omega_{f,\varphi} (\mathbf{Z},M,K) < r_s - 2\epsilon \right)  \leq \frac{1}{n} + \frac{1}{n\epsilon} + 2^{-n\epsilon}    
\end{equation}
proving that the code has typical authentication rate $r_s-2\epsilon$ with failure tolerance $ \frac{1}{n} + \frac{1}{n\epsilon} + 2^{-n\epsilon}.$

\end{IEEEproof}

\subsection{Proof of Theorem~\ref{thm:lai}} \label{app:proof_lai_2}
\begin{IEEEproof}
First note that there exists a sequence of $(0,r_n,\alpha_n,\epsilon_n,n)$-codes for the DM-BCCC$(p_{Y|X},p_{Z|X})$ such that 
\begin{align}
\lim_{n \rightarrow \infty}(r_n,\alpha_n,\epsilon_n) = (r,\alpha,0) 
\end{align}
for all $r$ and $\alpha$ 
\begin{align}
r+\alpha &\leq \mathbb{I}(Y;U,W) \\
\alpha &\leq \mathbb{I}(Y;U|W) - \mathbb{I}(Z;U|W)
\end{align}
by Corollary~\ref{cor:ck1713}.
Note, we may assume $\lim_{n\rightarrow \infty} n\epsilon = \infty$ since it is always possible to inject error into a decoder. 
This also implies a sequence of $(r_n,\alpha_n-2\epsilon_n, \alpha_n , \epsilon_n + \frac{1}{n} + \frac{1}{n\epsilon_n} + 2^{-n\epsilon_n},n)$-TA codes for the DM-AIC$(p_{Y|X},p_{Z|X})$ where  
\begin{align}
\lim_{n \rightarrow \infty}(r_n,\alpha_n-2\epsilon_n, \epsilon_n + \frac{1}{n} + \frac{1}{n\epsilon_n} + 2^{-n\epsilon_n}) = (r,\alpha,\alpha) , \label{eq:proof_lai_2:lp}
\end{align}
via Theorem~\ref{thm:pamtransform}.
Combining Equation~\eqref{eq:proof_lai_2:lp} with the operational definitions proves that if 
\begin{align}
r+\alpha &\leq \mathbb{I}(Y;U,W) \\
\alpha &\leq \mathbb{I}(Y;U|W) - \mathbb{I}(Z;U|W) \\
\alpha - \kappa &\leq 0 ,
\end{align}
then $(r,\alpha,\kappa) \in \mcf{C}_{\mathrm{TA}}(p_{Y|X},p_{Z|X})$.
\end{IEEEproof}

\section{Proof of Theorem~\ref{lem:-t} }\label{sec:-t}\label{app:cc2}

In order to prove Theorem~\ref{lem:-t}, it will be necessary to first prove the following theorem.
\begin{theorem}\label{thm:-tfinn}
For $n\geq 3$ and $\delta < \frac{5}{24},$ if a $(r,\alpha,\kappa,\delta,n)$ code for DM-AIC$(p_{Y|X},p_{Z|X})$ exists, then for each $\beta \in \left[2 \frac{\logt \gamma n}{n} ,r\right] $, where $\gamma \defn \frac{(4r+2)\ln 2}{-1+2\ln 2}$, there also exists a
$$(r-\beta,\alpha + \beta - \tilde \delta_n , \kappa + 2\beta, \tilde \delta_n ,n) \text{-code},$$
where $\tilde \delta_n \defn \max(2 \logt(\gamma n)/n,2\delta + \sqrt{\delta}+ (\gamma n)^{-1})  . $
\end{theorem}
Theorem~\ref{thm:-tfinn} is proved in Appendices~\ref{app:cc2:cc}--\ref{app:cc2:t1e}, and constitutes the majority of the work necessary in proving Theorem~\ref{lem:-t}.
Using Theorem~\ref{thm:-tfinn}, we prove Theorem~\ref{lem:-t} in Appendix~\ref{app:-t:proof}.
First, we describe the proof of Theorem~\ref{thm:-tfinn}.


The code construction engages by being given a $(r,\alpha,\kappa,\delta,n)$ code, $(f,\varphi)$, for DM-AIC$(p_{Y|X},p_{Z|X})$ and from it randomly selecting a new code $(\tilde f,\tilde \varphi)$. 
Here, the set of $2^{n(\kappa + 2\beta)}$ secret keys for $(\tilde f,\tilde \varphi)$ will be represented as two smaller secret keys chosen from sets of size $2^{n\kappa}$ and $2^{n2\beta}$. 
For each of the secret keys from the set of size $2^{n2\beta}$, there will exist an injective mapping from the set of messages for $(\tilde f, \tilde \varphi)$ ($2^{n(r-\beta)}$ elements) to the set of messages for $(f,\varphi)$ ($2^{nr}$ elements).

For transmission of a given message using $(\tilde f,\tilde \varphi)$, encoder $\tilde f$ acts by applying the injective mapping associated with the secret key from the set of size $2^{n2\beta}$ to the input message, and then using the resulting message and the secret key from the set of size $2^{n\kappa}$ as the input to the original encoder $f$. 
On the other end, the decoder $\tilde \varphi$ first applies the decoder $\varphi$ to the received message with appropriate secret key, and then inverts the injective mapping with the appropriate secret key.  
If the symbol can not be inverted, then deception is declared.  
Using this scheme, it is immediately clear that the resulting message rate of the code is $r-\beta$ and the resulting key consumption rate is $\kappa + 2\beta.$
This leaves the determination of the authentication rate and probability of error.

The authentication rate and probability of error of $(\tilde f,\tilde \varphi)$ will be put in terms of these same measures for $(f,\varphi)$.
In order to assist the preceding, let $\mbf{\tilde Z}, \tilde M,K_1,K_2$ be the RVs representing Gr{\'i}ma's observation, the message, first secret key, and second secret key of $(\tilde f,\tilde \varphi)$, respectively, while letting $\mbf{Z}, M,K$ be the RVs representing Gr{\'i}ma's observation, the message, and secret key of $(f,\varphi)$, respectively. 
Furthermore, let correlated RVs $(\tilde F,\tilde \Phi)$ represent the randomly chosen encoder and decoder.

In appendix~\ref{app:cc2:cc}, a code construction method is presented which specifies RVs $(\tilde F,\tilde \Phi)$. 
With regards to the message error of this code construction, 
\begin{align}
\Pr \left( \varepsilon_{\tilde F,\tilde \Phi} \geq \sqrt{\delta} \right) \leq \sqrt{\delta} 
\end{align}
is shown in Appendix~\ref{app:cc2:mea}.
While for the authentication rate analysis, it is shown in Appendix~\ref{app:cc2:t1e} that with probability greater than 
\begin{align}
1-2^{-2(n(r+1)-1)}
\end{align}
a code $(\tilde F,\tilde \Phi) = (\tilde f,\tilde \varphi)$ is chosen such that 
\begin{align}
\Pr\left( \omega_{\tilde f,\tilde \varphi}(\mbf{\tilde Z},\tilde M,K_1,K_2) \geq 2^{-n\left( \alpha + \beta - 2\logt (\gamma n)/n \right)} \right) \leq 2\delta+ \frac{1}{\gamma n},  \label{eq:cc2:pre:1}
\end{align}
where $\gamma = \frac{(4r+2) \ln 2}{-1 + 2\ln 2}.$

Note that there must exist at least one choice of $(\tilde f,\tilde \varphi)$ that satisfies Equation~\eqref{eq:cc2:pre:1} and $\varepsilon_{\tilde f,\tilde \varphi} < \sqrt{\delta}$ simultaneously since $\delta < \frac{5}{24}$ and $n \geq 3$ guarantees $1 - 2^{-(n(r+1)-1)} - 2\delta - \frac{1}{\gamma n} > 0$.
Hence, proving the existence of the $$(r-\beta,\alpha + \beta - 2 \logt(\gamma n)/n, \kappa + 2\beta, \sqrt{\delta} + 2^{-n(r+1)-1}+ 2\delta + (\gamma n)^{-1},n) \text{-code}$$
guaranteed in the theorem statement.

\begin{IEEEproof}

\subsection{Code Construction}\label{app:cc2:cc}

For a given positive real number $\beta \leq r$, we shall use the following construction to transform codes designed to send messages chosen uniformly from $\mcf{M}\defn \{ 1,\dots, 2^{nr}\}$ with a secret key drawn uniformly from $\mcf{K}_1 \defn \{ 1, \dots, 2^{n\kappa}\}$, into codes to send messages chosen uniformly from $\mcf{\tilde M} \defn \{ 1,\dots, 2^{n(r-\beta)} \}$ with a secret key drawn uniformly from $\mcf{K}_1 \times \mcf{K}_2,$ where $\mcf{K}_2 \defn \{ 1,\dots, 2^{n 2\beta} \}.$
The starting codes will be denoted $(f,\varphi) \in \mcf{P}(\mbcf{X}|\mcf{M}, \mcf{K}_1) \times \mcf{P}(\mcf{M} \cup \set{\mbf{!}} | \mbcf{Y}, \mcf{K}_1)$, and the resulting code after the transformation will be denoted $(\tilde f,\tilde \varphi) \in \mcf{P}(\mbcf{X}|\mcf{\tilde M}, \mcf{K}_1,\mcf{K}_2) \times \mcf{P}(\mcf{\tilde M} \cup \set{\mbf{!}} | \mbcf{Y}, \mcf{K}_1,\mcf{K}_2).$

\noindent  \textbf{Random codebook generation:} 
Independently for each $k_2 \in \mcf{K}_2$, select a mapping $g_{k_2}: \mcf{\tilde M} \rightarrow \mcf{M}$ uniformly from the set of all injective mappings from $\mcf{\tilde M}$ to $\mcf{M}$.

\noindent  \textbf{Encoders:}
\[
\tilde f(\mbf{x} | \tilde m,k_1,k_2) \defn f(\mbf{x} | g_{k_2}(\tilde m),k_1)
\]
for each $(\mbf{x},\tilde m,k_1,k_2) \in \mbcf{X} \times \mcf{\tilde M} \times \mcf{K}_1 \times \mcf{K}_2$.

\noindent  \textbf{Decoders:}
\[
\tilde \varphi(\tilde m|\mbf{y},k_1,k_2) = \begin{cases}
\varphi(g_{k_2}(\tilde m)|\mbf{y},k_1) &\text{ if } \tilde m \neq \mbf{!}  \\
\varphi(\mbf{!} | \mbf{y},k) + \varphi(\mcf{M} - g_{k_2}(\mcf{\tilde M})|\mbf{y},k_1) &\text{ otherwise} 
\end{cases},
\]
for all $(\mbf{y},k_1, k_2) \in \mbcf{Y}\times \mcf{K}_1 \times \mcf{K}_2$,  $\tilde m \in \mcf{\tilde M} \cup \mbf{!}$. 

\subsection{Message error analysis}\label{app:cc2:mea}
The average probability of message error over all possible $(\tilde f,\tilde \varphi)$ is equal to the probability of message error for $(f,\varphi)$. Indeed, this is a direct consequence of
\begin{align}
\varepsilon_{\tilde f,\tilde \varphi}(\tilde m,k_1,k_2) &= 1- \sum_{\subalign{\mbf{y}&\in \mbcf{Y}, \\ \mbf{x}&\in \mbcf{X} }} \tilde \varphi(\tilde m|\mbf{y},k_1,k_2)  p_{Y|X}(\mbf{y}|\mbf{x}) \tilde f(\mbf{x}|\tilde m,k_1,k_2) \notag \\
&= 1- \sum_{\subalign{\mbf{y}&\in \mbcf{Y}, \\ \mbf{x}&\in \mbcf{X} }}  \varphi(g_{k_2}(\tilde m)|\mbf{y},k_1)  p_{Y|X}(\mbf{y}|\mbf{x}) f(\mbf{x}|g_{k_2}(\tilde m),k_1) \notag \\
&= \varepsilon_{f,\varphi}(g_{k_2}(\tilde m),k_1),
\end{align}
and the fact that the mapping $g_{k_2}$ is chosen uniformly from the set of of all injective mappings.
Therefore,
\begin{equation}\label{eq:app:cc2:me1}
\mathbb{E}[ \varepsilon_{\tilde F,\tilde \Phi} ]  = \sum_{\tilde m \in \mcf{\tilde M}, k_1 \in \mcf{K}_1 } 2^{-n(r - \beta + \kappa)} \left( \sum_{m \in \mcf{M}} 2^{-n r } \varepsilon_{f,\varphi}(m,k_1)\right) = \varepsilon_{f,\varphi}  \leq \delta 
\end{equation}
since $g_{k_2}$ is chosen uniformly from the set of all injective mappings $\mcf{\tilde M} \rightarrow \mcf{M}$. 
Now, $$\Pr \left( \varepsilon_{\tilde F,\tilde \Phi} \geq \sqrt{\delta} \right) \leq \sqrt{\delta}$$
directly follows from combining Equation~\eqref{eq:app:cc2:me1} and Markov's inequality.

\subsection{Typical authentication rate analysis}\label{app:cc2:t1e}

Here, we shall show that  
\begin{equation}\label{eq:thm:-t:tt1e}
\Pr\left( \omega_{\tilde f,\tilde \varphi}(\mbf{Z},\tilde M,K_1,K_2) \geq 2^{-n\left( \alpha + \beta - 2\logt (\gamma n)/n \right)} \right) \leq 2\delta+ \frac{1}{\gamma n}
\end{equation}
as long as $(\tilde f,\tilde \varphi) \in \mcf{G}^* \cap \mcf{G}^\dagger$, where $\mcf{G}^*$ is the set of $(\tilde f,\tilde \varphi)$ for which
\begin{equation}\label{eq:thm:-t:suff2}
\abs{\set{k_2 : \{m,m'\} \subseteq g_{k_2}(\mcf{\tilde M}) }} \leq n\gamma ,
\end{equation}
for all $ m  \in \mcf{M}$ and $m' (\neq m) \in \mcf{M}$,
while $\mcf{G}^\dagger$ is the set of $(\tilde f,\tilde \varphi)$ such that
\begin{align}
\abs{\set{k_2 : m \in g_{k_2}(\mcf{\tilde M}) }} \leq 2^{1+n\beta}
\end{align}
for all $m \in \mcf{M}.$
For clarity of presentation, in Appendix~\ref{app:sims:ta}, we show that if $(\tilde f,\tilde \varphi) \in \mcf{G}^\dagger \cap \mcf{G}^*$, then Equation~\eqref{eq:thm:-t:tt1e} holds, and thus if $(\tilde f,\tilde \varphi) \in \mcf{G}^\dagger \cap \mcf{G}^*$, then the new code has an authentication rate of $\alpha + \beta - \logt(\gamma n)/n$ and an authentication tolerance of $2\delta + \frac{1}{\gamma n}$.
Next, in Appendix~\ref{app:sims:g*}, we show  
\begin{equation}\label{eq:sims:pg*}
\Pr \left( (\tilde F,\tilde \Phi) \notin \mcf{G}^* \right) \leq e^{-1} \cdot 2^{-2\left( n(r+1)-1 \right) } ,
\end{equation}
and in Appendix~\ref{app:sims:gdagger}, we show 
\begin{equation}\label{eq:sims:pgdagger}
\Pr \left( (\tilde F,\tilde \Phi) \notin \mcf{G}^\dagger \right) \leq  2^{nr}  e^{-(2\ln 2 - 1)  2^{n\beta}} 
\end{equation}
Thus if $\beta \geq \logt (\gamma n)/n$, then 
\begin{equation}\label{eq:sims:pg*dagger}
\Pr \left( (\tilde F,\tilde \Phi) \notin \mcf{G}^\dagger \cap \mcf{G}^* \right) \leq  2^{-2\left( n(r+1)-1 \right) }
\end{equation}
follows from Equations~\eqref{eq:sims:pg*} and~\eqref{eq:sims:pgdagger} and the union bound.
Since $2^{-2\left( n(r+1)-1 \right) }< 1$ for all $n\geq 2$, it proves there must exist at least one $(\tilde f,\tilde \varphi) \in \mcf{G}^\dagger \cap \mcf{G}^*$, thus proving the theorem statement.

Proofs of both Equations~\eqref{eq:sims:pg*} and~\eqref{eq:sims:pgdagger} will use the following lemma from Csisz{\'a}r and K{\"o}rner.
\begin{lemma}\label{lem:ck} (\cite[Lemma~17.9]{CK}) The probability that in $k$ independent trials an event of probability $q$ occurs less/more than $\alpha q k$ times, according as $\alpha \lessgtr 1$, is bounded above by $e^{-c(\alpha)qk}$ where $c(\alpha) = \alpha \ln \alpha - \alpha + 1$.
\end{lemma}

\begin{remark} This result implies that if $W_1,\dots,W_n$ are independent Bernoulli random variables and $t \geq \sum_{i=1}^n \mathbb{E}[W_i]$, then
$$\Pr \left( \sum_{i=1}^n W_i > t+ \sum_{i=1}^n \mathbb{E}[W_i]  \right) \leq \max_{u \in [0,t]} e^{-\left[(u+t)\ln \left(1 + \frac{t}{u} \right) - t\right]  } = e^{-(2\ln 2 -1 ) t }.$$
\end{remark}

\subsubsection{Typical authentication rate given $(\tilde f,\tilde \varphi) \in \mcf{G}^* \cap \mcf{G}^\dagger$} \label{app:sims:ta}

In proving Equation~\eqref{eq:thm:-t:tt1e} holds for all $(\tilde f,\tilde \varphi) \in \mcf{G}^* \cap \mcf{G}^\dagger$, it will be helpful to first prove that 
\begin{equation}\label{eq:app:-tfinn:auth:a:1}
p_{\mbf{\tilde Z},\tilde M,K_1,K_2}(\mbf{ z} ,  \tilde m, k_1,k_2) = p_{\mbf{ Z}, M,K}(\mbf{z},g_{k_2}(\tilde m),k_1) 2^{-n\beta},
\end{equation}
as well as prove that if $(\tilde f, \tilde \varphi) \in \mcf{G}^*$, then 
\begin{equation}\label{eq:app:-tfinn:auth:a:p2}
|\mcf{K}_2^*(\mbf{z},m,k)| < 2^{n\beta} (\gamma n)^{-1}
\end{equation}
where\footnote{Here $\frac{\omega_{\tilde f, \tilde \varphi}(\mbf{z},g_{k_2}^{-1}(m), k_1,k_2)}{\omega_{f,\varphi}(\mbf{z},m,k_1)   } \defn 0$ if $g_{k_2}^{-1}(m)$ does not exist.}
$$ \mcf{K}_2^*(\mbf{z},m,k) \defn \set{k_2 : \frac{\omega_{\tilde f, \tilde \varphi}(\mbf{z},g_{k_2}^{-1}(m), k_1,k_2)}{\omega_{f,\varphi}(\mbf{z},m,k_1)   } >  2^{-n\beta} (\gamma n)^2 }$$
for all $m \in \mcf{M}$, $\mbf{z} \in \mbcf{Z}, k \in \mcf{K}_1$.

First, Equation~\eqref{eq:app:-tfinn:auth:a:1} is a consequence of $\mbf{\tilde X}|\{\tilde M = \tilde m, K_1=k_1,K_2=k_2\} $ being the same as $\mbf{X}|\{M = g_{k_2}(\tilde m),K = k_1\}$ when $\tilde M, M, K_1,K_2,K$ are uniform over their support sets, since then
\begin{align*}
p_{\mbf{\tilde Z},\tilde M,K_1,K_2}(\mbf{ z} ,  \tilde m, k_1,k_2) &= p_{\mbf{\tilde Z}| \tilde M,K_1,K_2}(\mbf{z}|\tilde m,k_1,k_2) 2^{-n(r + \beta + \kappa)} \\
&= p_{\mbf{ Z}|M,K}(\mbf{z}|g_{k_2}(\tilde m),k_1) 2^{-n(r + \beta + \kappa)} \\
&= p_{\mbf{ Z},M,K}(\mbf{z},g_{k_2}(\tilde m),k_1) 2^{-n\beta}
\end{align*}
follows. 

Next, for all $\mbf{z},m,k_1$, Equation~\eqref{eq:app:-tfinn:auth:a:p2} can be derived as follows: 
\begin{align}
&|\mcf{K}_2^*(\mbf{z},m,k_1)| 2^{-n\beta}(\gamma n)^2 \omega_{f,\varphi}(\mbf{z},m,k_1)\notag \\
&\quad < \sum_{k_2 \in \mcf{K}_2^*(\mbf{z},m,k_1)}  \omega_{\tilde f,\tilde \varphi}(\mbf{z},g_{k_2}^{-1}(m),k_1,k_2) \label{eq:app:-tfinn:auth:a:p2:1}\\
&\quad = \sum_{k_2 \in \mcf{K}_2^*(\mbf{z},m,k_1)} \sum_{\mbf{y},\tilde m' \neq g_{k_2}^{-1}(m)} \psi(\mbf{y}|\mbf{z}) \tilde \varphi(\tilde m'| \mbf{y},k_1,k_2) \label{eq:app:-tfinn:auth:a:p2:2} \\
&\quad = \sum_{k_2 \in \mcf{K}_2^*(\mbf{z},m,k_1)} \sum_{\mbf{y},m'\neq m} \psi(\mbf{y}|\mbf{z})  \varphi(m' | \mbf{y},k_1) \idc{m' \in g_{k_2}(\mcf{\tilde M}) }  \label{eq:app:-tfinn:auth:a:p2:3} \\
&\quad \leq \sum_{\mbf{y},m'\neq m} \psi(\mbf{y}|\mbf{z})  \varphi(m' | \mbf{y},k_1)  \sum_{k_2} \idc{\{ m,m'\} \subset   g_{k_2}(\mcf{\tilde M}) }  \label{eq:app:-tfinn:auth:a:p2:4} \\
&\quad \leq (\gamma n) \omega_{f,\varphi}(\mbf{z},m,k_1)  \label{eq:app:-tfinn:auth:a:p2:5};
\end{align}
where~\eqref{eq:app:-tfinn:auth:a:p2:1} is because $$|\mcf{K}_2^*(\mbf{z},m,k_1)| 2^{-n\beta}(\gamma n)^2 \omega_{f,\varphi}(\mbf{z},m,k_1)  = \sum_{k_2 \in \mcf{K}_2^*(\mbf{z},m,k_1)} 2^{-n\beta}(\gamma n)^2 \omega_{f,\varphi}(\mbf{z},m,k_1) $$
and $2^{-n\beta} (\gamma n)^2 \omega_{f,\varphi}(\mbf{z},m,k_1)  \leq \omega_{\tilde f,\tilde \varphi}(\mbf{z},g_{k_2}^{-1}(m),k_1,k_2)$ for all $k_2 \in \mcf{K}_2^*(\mbf{z},m,k_1);$~\eqref{eq:app:-tfinn:auth:a:p2:2} is by the definition of function $\omega;$~\eqref{eq:app:-tfinn:auth:a:p2:3} is by the definition of function $\tilde \varphi;$~\eqref{eq:app:-tfinn:auth:a:p2:4} is by exchanging the summation basis and recognizing that $\mcf{K}_2^*(\mbf{z},m,k_1) \subseteq \{ k_2 : m \in g_{k_2}(\mcf{\tilde M})\}$; finally~\eqref{eq:app:-tfinn:auth:a:p2:5} is because $(\tilde f, \tilde \varphi) \in \mcf{G}^*$ and by the definition of $\omega.$

With Equations~\eqref{eq:app:-tfinn:auth:a:1} and~\eqref{eq:app:-tfinn:auth:a:p2} in hand, the probability that $\omega_{\tilde f,\tilde \varphi}(\mbf{\tilde Z},\tilde M,K_1,K_2) \geq 2^{-n(\alpha + \beta - 2 \logt (\gamma n) /n}$ can be upper bounded by putting it in terms of the probability that $\omega_{f, \varphi}(\mbf{ Z}, M,K) \geq 2^{-n\alpha }$ as follows:
\begin{align}
&\sum_{\substack{\mbf{z},\tilde m,k_1,k_2:\\ \omega_{\tilde f,\tilde \varphi}(\mbf{z},\tilde m,k_1,k_2) \geq 2^{-n(\alpha+\beta-2\logt(\gamma n)/n )}  } } p_{\mbf{\tilde Z},\tilde M,K_1,K_2}(\mbf{ z} ,  \tilde m, k_1,k_2) \notag  \\
&=\sum_{\substack{\mbf{z},m,\tilde m,k_1,k_2:\\ \omega_{\tilde f,\tilde \varphi}(\mbf{z},g_{k_2}^{-1}(m),k_1,k_2) \geq 2^{-n(\alpha+\beta-2 \logt (\gamma n)/n)}  } } \idc{g_{k_2}(\tilde m) = m} p_{\mbf{\tilde Z},\tilde M,K_1,K_2}(\mbf{ z} ,  \tilde m, k_1,k_2)  \label{eq:app:-tfinn:auth:a:2}\\
&=\sum_{\substack{\mbf{z},m,\tilde m,k_1,k_2:\\ \omega_{\tilde f,\tilde \varphi}(\mbf{z},g_{k_2}^{-1}(m),k_1,k_2) \geq 2^{-n(\alpha+\beta-2\logt(\gamma n)/n )}  } } \idc{g_{k_2}(\tilde m) = m} p_{\mbf{Z}, M,K}(\mbf{ z} , m, k_1)  2^{-n\beta} \label{eq:app:-tfinn:auth:a:3}\\
&=\sum_{\substack{\mbf{z},m,k_1,k_2:\\ \omega_{\tilde f,\tilde \varphi}(\mbf{z},g_{k_2}^{-1}( m),k_1,k_2) \geq 2^{-n(\alpha+\beta-2 \logt(\gamma n) /n )}  } } \idc{m \in g_{k_2}(\mcf{\tilde M}) } p_{\mbf{Z}, M,K}(\mbf{ z} , m, k_1)  2^{-n\beta} \label{eq:app:-tfinn:auth:a:4}\\
&\leq \sum_{\substack{\mbf{z},m,k_1,\\ k_2 \in \mcf{K}_2^*(\mbf{z},m,k_1)  } } \idc{m \in g_{k_2}(\mcf{\tilde M}) } p_{\mbf{Z}, M,K}(\mbf{ z} , m, k_1)  2^{-n\beta} \notag \\
&\quad + \sum_{\substack{\mbf{z},m,k_1,k_2:\\ \omega_{ f, \varphi}(\mbf{z},m,k_1) \geq 2^{-n\alpha}  } } \idc{m \in g_{k_2}(\mcf{\tilde M}) } p_{\mbf{Z}, M,K}(\mbf{ z} , m, k_1)  2^{-n\beta} \label{eq:app:-tfinn:auth:a:5}\\
&\leq \sum_{\mbf{z},m,k_1 }  p_{\mbf{Z}, M,K}(\mbf{ z} , m, k_1)  (\gamma n)^{-1}  + \sum_{\substack{\mbf{z},m,k_1:\\ \omega_{ f, \varphi}(\mbf{z},m,k_1) \geq 2^{-n\alpha}  } }  p_{\mbf{Z}, M,K}(\mbf{ z} , m, k_1)  2 \label{eq:app:-tfinn:auth:a:6}\\
&\leq  (\gamma n)^{-1}  + 2\delta  \label{eq:app:-tfinn:auth:a:7}
\end{align}
where~\eqref{eq:app:-tfinn:auth:a:2} is because $g_{k_2}$ is an injective mapping and thus the is a single $m$ for each $k_2,\tilde m;$~\eqref{eq:app:-tfinn:auth:a:3} is by the earlier observation of~\eqref{eq:app:-tfinn:auth:a:1};~\eqref{eq:app:-tfinn:auth:a:4} follows by summing over $\tilde m \in \mcf{\tilde M};$~\eqref{eq:app:-tfinn:auth:a:5} follows by splitting the summation terms based upon whether or not $\omega_{\tilde f, \tilde \varphi}(\mbf{z},g_{k_2}^{-1}(m), k_1,k_2)   >  2^{-n\beta}(\gamma n)^{2}\omega_{f,\varphi}(\mbf{z},m,k_1) $, and then recognizing that $\omega_{ f, \varphi}(\mbf{z},m,k_1) \geq 2^{-n\alpha}$ for all $(\mbf{z},m,k_1,k_2)$ such that $\omega_{\tilde f, \tilde \varphi}(\mbf{z},g_{k_2}^{-1}(m), k_1,k_2) \leq  2^{-n\beta}(\gamma n)^{2}\omega_{f,\varphi}(\mbf{z},m,k_1)$ and $\omega_{\tilde f,\tilde \varphi}(\mbf{z},g_{k_2}^{-1}( m),k_1,k_2) \geq 2^{-n(\alpha+\beta-2\logt(\gamma n)/n)}$;~\eqref{eq:app:-tfinn:auth:a:6} is because $(\tilde f,\tilde \varphi) \in \mcf{G}^*\cap \mcf{G}^\dagger$, where more specifically the first summation's bound is due to Equation~\eqref{eq:app:-tfinn:auth:a:p2} since $(\tilde f, \tilde \varphi) \in \mcf{G}^*$ and the second summation's bound is since $(\tilde f,\tilde \varphi) \in \mcf{G}^\dagger;$ and finally~\eqref{eq:app:-tfinn:auth:a:7} is by the law of total probability and because $(f,\varphi)$ is assumed to be a $(r,\alpha,\kappa,\delta,n)$-TA code. 
Equation~\eqref{eq:app:-tfinn:auth:a:7} confirms Equation~\eqref{eq:thm:-t:tt1e}.

\subsubsection{Probability $(\tilde F,\tilde \Phi) \in \mcf{G}^*$}\label{app:sims:g*}

To prove Equation~\eqref{eq:sims:pg*}, first fix any $m \in \mcf{M}$ and $m' (\neq m) \in \mcf{M}$ and let
\begin{align}
A_{k_2} \defn \idc{ \{m,m'\} \subseteq G_{k_2}(\mcf{\tilde M}) }  \label{eq:sims:ar:0}.
\end{align}
Of importance is that $A_{k_2}$ and $A_{k_2'}$ are independent for $k_2 \neq k_2'$ since the mappings $G_{k_2}$ are independently chosen for each $k_2 \in \mcf{K}_2$.
Furthermore,
\begin{align}
\Pr \left(A_{k_2}=1 \right) &=\left.\left( \begin{matrix} 2^{nr}-2 \\ 2^{n(r-\beta)}-2 \end{matrix} \right)\middle/ \left( \begin{matrix} 2^{nr} \\ 2^{n(r-\beta)} \end{matrix} \right) \right. 
\notag \\ &
=2^{-n\beta} \frac{2^{n(r-\beta)}-1}{2^{nr} - 1} 
= 2^{-2n \beta} \frac{1 - 2^{n(\beta - r) }}{1 - 2^{-nr}} \leq 2^{-2n\beta} \label{eq:app:cc2:t1e:nts3}
\end{align}
since $G_{k_2}(\mcf{\tilde M})$ is uniform over the size $2^{n(r-\beta)}$ subsets of $\mcf{M}.$
As a result of these properties,
\begin{align}
\Pr \left( \sum_{k_2 \in \mcf{K}_2} A_{k_2}  >  \gamma n  \right) &\leq \Pr \left( \sum_{k_2 \in \mcf{K}_2} A_{k_2}  >  \gamma n - 1 + \sum_{k_2 \in \mcf{K}_2} \mathbb{E}[A_{k_2}] \right) \notag \\
&\leq e^{-(2\ln 2 - 1) (  \gamma n -1) } = e^{-1} \cdot 2^{-4nr - 2(n-1)} \label{eq:ar:sims:finaleq1}
\end{align}
follows by applying Lemma~\ref{lem:ck}.
Equation~\eqref{eq:sims:pg*} now follows by using the union bound to consider all $m \in \mcf{M}$ and $m' (\neq m) \in \mcf{M}$ simultaneously. 

\subsubsection{Probability $(\tilde F,\tilde \Phi) \in \mcf{G}^\dagger$}\label{app:sims:gdagger}

First, fix a $m \in \mcf{M}$, and let $B_{k_2} = \idc{m \in G_{k_2}(\mcf{\tilde M})}.$
Clearly, $B_{k_2}$ and $B_{k_2'}$ are independent for $k_2 \neq k_2'$ since $G_{k_2}$ and $G_{k_2'}$ are independent. 
Furthermore, $\Pr \left( B_{k_2} =1 \right) = 2^{-n\beta}$ since $G_{k_2}(\mcf{\tilde M})$ is uniform over the size $2^{n(r-\beta)}$ subsets of $\mcf{M}$.
Hence,
\begin{align}
\Pr \left( \sum_{k_2 \in \mcf{K}_2} B_{k_2} - 2^{n\beta} > 2^{n\beta}  \right) &\leq   e^{-(2\ln 2 - 1)  2^{n\beta}}   \label{eq:ar:sims:finaleq2}
\end{align}
follows from Lemma~\ref{lem:ck} since 
$$ \mathbb{E}\left[ \sum_{k_2 \in \mcf{K}_2} B_{k_2} \right] =\sum_{k_2 \in \mcf{K}_2} 2^{-n\beta} = 2^{n\beta}. $$
Equation~\eqref{eq:sims:pgdagger} now follows by using the union bound to consider all $m \in \mcf{M}$ simultaneously. 

\end{IEEEproof}

\subsection{Proof of Theorem~\ref{lem:-t}}\label{app:-t:proof}

\begin{IEEEproof}

If $(r,\alpha,\kappa) \in \mcf{C}_{\mathrm{TA}}(p_{Y|X},p_{Z|X})$, then there exists a sequence of $(r_n,\alpha_n,\kappa_n,\delta_n,n)$ codes, $(f_n,\varphi_n)$, such that 
\begin{align}
\lim_{n\rightarrow \infty} |(r_n,\alpha_n,\kappa_n,\delta_n) - (r,\alpha,\kappa,0) |= 0.
\end{align}
By definition, then, there must exist an $n'\geq 3$ such that $\delta_n \leq \frac{5}{24}$ and $r_n \geq 2 \frac{\logt \gamma n}{n}$ $\left(\text{where } \gamma = \frac{(4r+2)\ln 2}{-1+2\ln 2} \right)$ for all $n\geq n'.$
Hence, for any positive $\beta < r$, there also exists a sequence of 
$$(r_n-\beta_n , \alpha_n + \beta_n - 2\frac{\logt \gamma n}{n}, \kappa_n + 2\beta_n, \sqrt{\delta_n} + 2^{-n(r+1)-1}+ 2\delta_n + \frac{1}{\gamma n}, n)\text{ codes},$$
where $\lim_{n\rightarrow \infty} \beta_n = \beta$, by Theorem~\ref{thm:-tfinn}.
Taking the limit point of this sequence of codes proves that $$(r-\beta,\alpha+\beta,\kappa + 2\beta) \in \mcf{C}_{\mathrm{TA}}(p_{Y|X},p_{Z|X}).$$

\end{IEEEproof}

\section{Proof of Lemma~\ref{lem:c:2}}\label{app:c}

We shall break the proof of Lemma~\ref{lem:c:2} into two parts.
In both parts, we shall assume that $(f,\varphi)$ is a $(r,\alpha,\kappa,\delta,n)$-TA code for DM-AIC$(p_{Y|X},p_{Z|X})$, and then show in Appendix~\ref{app:c:2:1} that this requires
\begin{equation}\label{eq:lem:c:eq:1}
\alpha \leq n^{-1}\mathbb{I}(\mbf{Y};K) + \zeta(\delta,n)
\end{equation}
for some $\zeta(\delta,n)$ such that $\lim_{\subalign{\delta &\rightarrow 0^+\\ n & \rightarrow \infty}} \zeta(\delta,n) = 0$, and similarly show in Appendix~\ref{app:c:2:2} it also requires
\begin{equation}\label{eq:lem:c:eq:2}
\alpha \leq n^{-1}\mathbb{H}(K|\mbf{Z}) + \tilde \zeta(\delta,n)
\end{equation}
for some $\tilde \zeta(\delta,n)$ such that $\lim_{\subalign{\delta &\rightarrow 0^+\\ n & \rightarrow \infty}} \tilde \zeta(\delta,n) = 0$.
Clearly, having validated Equations~\eqref{eq:lem:c:eq:1} and~\eqref{eq:lem:c:eq:2}, then Lemma~\ref{lem:c:2} will follow by choosing the larger of $\tilde \zeta$ and $\zeta$ as the function presented in the lemma statement. 

For the proofs of both Equations~\eqref{eq:lem:c:eq:1} and~\eqref{eq:lem:c:eq:2}, we shall make use of the partitioning random variable, $T$, constructed in~\cite{graves2018inducing}. 
Discussion on the properties of the random variable can be found in Section~\ref{sec:inducing}.
For these proofs we shall use the sequence $\lambda_n$ discussed prior, for which $\lim_{n\rightarrow \infty} \lambda_n = 0$ and $\lim_{n\rightarrow \infty} n \lambda_n = \infty.$ 
We shall also introduce a new sequence of error terms, $\nu_n \defn \delta + 3 \cdot 2^{-n\lambda_n},$ which converge to $\delta$ as $n\rightarrow \infty$, and furthermore converge to $0$ if $n\rightarrow \infty$ and $\delta \rightarrow 0.$

Before moving to proving Equations~\eqref{eq:lem:c:eq:1} and~\eqref{eq:lem:c:eq:2}, we will need to prove the following technical lemma. 
\begin{lemma}\label{lem:pre:c:2}
For $ r \geq 2 \lambda_n$, if $(f,\varphi)$ is a $(r,\alpha,\kappa,\delta,n)$-TA code for DM-AIC$(p_{Y|X},p_{Z|X})$, then
$$ \Pr \left( \tau(M,K,T) < 1 - \sqrt{\nu_n }\right) \leq \sqrt{\nu_n}  $$
where
\begin{align*}
\tau(m,k,t) &\defn \sum_{\mbf{y} : (\mbf{y},k,t) \in \mcf{D}^+ } p(\mbf{y}|k,t)  \varphi(\mcf{M}-\{m\}|\mbf{y},k),
\end{align*} 
and $\nu_n \defn \delta + 3\cdot 2^{-n\lambda_n}.$

\end{lemma}

\begin{IEEEproof}
First, observe the following lower bounds on the expectation of $\tau(M,K,T);$
\begin{align}
&\sum_{m,k,t} p(m,k,t) \tau(m,k,t) \notag \\
&\quad \geq -2^{-n\lambda_n} + \sum_{m,k,t} p(m,k,t) \sum_{\mbf{y}} p(\mbf{y}|k,t) \varphi(\mcf{M}-\{m\}|\mbf{y},k) \label{eq:pre:c:2:0} \\
&\quad \geq -2^{-n\lambda_n} +  \sum_{m,k,t} -p(m,k,t)p(m|k,t) + \sum_{\mbf{y},m,k,t} p(\mbf{y},m,k,t) \varphi(m|\mbf{y},k) \label{eq:pre:c:2:1}\\
&\quad \geq -2^{-nr + n\lambda_n} - 2\cdot 2^{-n\lambda_n} + 1 - \delta  \geq 1 - \nu_n \label{eq:pre:c:2:2};
\end{align}
where~\eqref{eq:pre:c:2:0} follows because the probability that $(\mbf{Y},M,K,T) \notin \mcf{D}^+$ is less than $2^{-n\lambda_n};$~\eqref{eq:pre:c:2:1} is because
\begin{align*}
\sum_{\mbf{y}} p(\mbf{y}|k,t) \varphi(\mcf{M}-\{m'\}|\mbf{y},k) & \geq \sum_{\mbf{y},m} p(\mbf{y},m|k,t) \varphi(m-\{m'\}|\mbf{y},k)\\
&\geq -p_{M|K,T}(m'|k,t) + \sum_{\mbf{y},m} p(\mbf{y},m|k,t) \varphi(m|\mbf{y},k);
\end{align*}
and~\eqref{eq:pre:c:2:2} is because the probability $p(M|K,T) < 2^{-nr + n\lambda_n}$ is less than $2^{-n\lambda_n}$, the probability of message error must be less than $1-\delta,$ and $r > 2 \lambda_n.$
Next, observe the following upper bound on the expectation of $\tau(M,K,T);$
\begin{align}
&\sum_{m,k,t} p(m,k,t) \tau(m,k,t) \notag \\
&\quad \leq \sum_{\substack{m,k,t:\\ \tau(m,k,t) > 1-\sqrt{\nu_n} }} p(m,k,t) + \sum_{\substack{m,k,t:\\ \tau(m,k,t) \leq  1-\sqrt{\nu_n } }} p(m,k,t)\left( 1-\sqrt{\nu_n }\right) \label{eq:pre:c:2:3} \\ 
&\quad = 1 - \Pr \left( \tau(M,K,T) \leq  1- \sqrt{\nu_n} \right) \left(1- \sqrt{\nu_n} \right).
\end{align}
Combining these two observations and solving for $\Pr \left( \tau(M,K,T) \leq  1- \sqrt{\nu_n} \right)$ proves the lemma statement.

\end{IEEEproof}

With this technical lemma in hand, we proceed to the proof of Lemma~\ref{lem:c:2}.

\begin{IEEEproof}

\subsection{$\alpha \leq n^{-1}\mathbb{I}(\mbf{Y};K) + \zeta(\delta,n)$}\label{app:c:2:1}
Let $(f,\varphi)$ be a $(r,\alpha,\kappa,\delta,n)$-TA code for DM-AIC$(p_{Y|X},p_{Z|X})$. 
By definition, a $(r,\alpha,\kappa,\delta,n)$-TA code requires
\begin{align}
\delta &\geq \max_{\psi \in \mbcf{P}(\mbcf{Y}|\mbcf{Z})} \Pr \left(  \sum_{\mbf{y}} \psi(\mbf{y}|\mbf{Z})  \varphi(\mcf{M}-\{M\}|\mbf{y},K) \geq 2^{-n\alpha}  \right)  \\
&\geq  \Pr \left(  \sum_{\mbf{y}} p_{\mbf{Y}}(\mbf{y})  \varphi(\mcf{M}-\{M\}|\mbf{y},K) \geq 2^{-n\alpha}  \right)  \label{eq:lem:c:2:codeass}
\end{align}
since $p_{\mbf{Y}} \in \mbcf{P}(\mbcf{Y}|\mbcf{Z})$.
Now, introducing the information stabilizing random variable $T$ (see Section~\ref{sec:inducing}) into the RHS of~\eqref{eq:lem:c:2:codeass} provides
\begin{align}
\delta &\geq  \sum_{m,k,t}  \Pr \left(  \sum_{ \tilde t\in \mcf{T}} p_{T}(\tilde t) \sum_{\mbf{y} } p(\mbf{y}|\tilde t)  \varphi(\mcf{M}-\{m\}|\mbf{y},k) \geq 2^{-n\alpha}  \right)  p(m,k,t)\\
&\geq  \sum_{m,k,t}  \idc{  p(t) \sum_{\mbf{y} \in \mbcf{Y}: (\mbf{y},k,t) \in \mcf{D}^+} p(\mbf{y}| t)  \varphi(\mcf{M}-\{m\}|\mbf{y},k) \geq 2^{-n\alpha}  }  p(m,k,t)\label{eq:lem:c:2:withs},
\end{align}
where $\mcf{D}^+$ is the stabilized set discussed in Section~\ref{sec:inducing}, since all summands inside the indicator are positive.
But,
\begin{align}
p(\mbf{y}|t) = \frac{p(\mbf{y}|t)}{p(\mbf{y}|k,t)} p(\mbf{y}|k,t) \geq 2^{-\mathbb{I}(\mbf{Y};K|T=t) - 2n\lambda_n} p(\mbf{y}|k,t)\label{eq:lem:c:2:withs2}
\end{align}
for all $(\mbf{y},k,t) \in \mcf{D}^+.$
Hence, 
\begin{align}
\delta &\geq     \sum_{m,k,t}   \idc{  2^{-\mathbb{I}(\mbf{Y};K|T=t)-2n\lambda_n}  p(t) \tau(m,k,t)  \geq 2^{-n\alpha}  } p(m,k,t) \label{eq:lem:c:2:ssetup}
\end{align}
by combining Equations~\eqref{eq:lem:c:2:withs} and~\eqref{eq:lem:c:2:withs2}. 
Furthermore,
\begin{align}
\delta + \sqrt{\nu_n} &\geq     \sum_{t}   \idc{  2^{-\mathbb{I}(\mbf{Y};K|T=t)-2n\lambda_n}  p(t) \left(1-\sqrt{\nu_n}\right)  \geq 2^{-n\alpha}  } p(t) \label{eq:lem:c:2:ssetup2},
\end{align}
recalling that $\nu_n \defn \delta + 3\cdot 2^{-n\lambda_n}$,
by using that
\begin{align*}
&  \idc{  2^{-\mathbb{I}(\mbf{Y};K|T=t)-2n\lambda_n}  p(t) \tau(m,k,t)  \geq 2^{-n\alpha}  } \\
&\quad\geq  \idc{  2^{-\mathbb{I}(\mbf{Y};K|T=t)-2n\lambda_n}  p(t)\left(1-\sqrt{\nu_n}\right)  \geq 2^{-n\alpha}  } - \idc{  \tau(m,k,t) < 1 - \sqrt{\nu_n}}
\end{align*}
and then applying Lemma~\ref{lem:pre:c:2} to the sum of the $\idc{  \tau(m,k,t) < 1 - \sqrt{\nu_n}}$ terms. 

Equation~\eqref{eq:lem:c:2:ssetup2} provides a bound on the number of $t$ for which $\mathbb{I}(\mbf{Y};K|T=t)$ can be less than $n\alpha$.
With this in mind, observe that
\begin{align}
&\mathbb{I}(\mbf{Y};K) + 2 \mathbb{H}(T) + 2n\lambda_n -\logt\left(1-\sqrt{\nu_n} \right) \notag \\
&\quad \geq \sum_{t } p(t) \left[ \mathbb{I}(\mbf{Y};K|T=t) - \logt p(t) + 2n\lambda_n -\logt\left(1-\sqrt{\nu_n}\right)\right]  \\
&\quad \geq n\alpha \sum_{t} p(t)  \idc{ 2^{- \mathbb{I}(\mbf{Y};K|T=t)-2n\nu_n }p(t)(1-\sqrt{\nu_n}) < 2^{-n\alpha}}  \\
&\quad \geq n\alpha( 1 - \delta - \sqrt{\nu_n}), \label{eq:lem:c:2:fin1} 
\end{align}
where~\eqref{eq:lem:c:2:fin1} is where Equation~\eqref{eq:lem:c:2:ssetup2} is specifically used. 
Collecting all the vanishing terms in Equation~\eqref{eq:lem:c:2:fin1} yields 
\begin{align}
n^{-1}\mathbb{I}(\mbf{Y};K) + \zeta(\delta,n) & \geq \alpha \label{eq:lem:c:2:fin} 
\end{align}
where
$$\zeta(\delta,n) = \frac{ (\delta+ \sqrt{\nu_n}) \logt |\mcf{Y}| + 4 \lambda_n  - n^{-1} \logt (1-\sqrt{\nu_n})}{1 - \delta - \sqrt{\nu_n}},$$
since $\mathbb{I}(\mbf{Y};K) \leq \logt |\mcf{Y}|$ and $\mathbb{H}(T) \leq \logt |\mcf{T}| \leq n \lambda_n.$
This proves Equation~\eqref{eq:lem:c:eq:1} since $\lim_{\subalign{\delta &\rightarrow 0^+,\\ n&\rightarrow \infty}} \zeta(\delta,n) = 0  .$

\subsection{$\alpha \leq n^{-1} \mathbb{H}(K|\mbf{Z}) + \zeta(\delta,n)$}\label{app:c:2:2}

Once again, let $(f,\varphi)$ be a $(r,\alpha,\kappa,\delta,n)$-TA code for DM-AIC$(p_{Y|X},p_{Z|X})$. 
Being a $(r,\alpha,\kappa,\delta,n)$-TA code requires that
\begin{align}
\delta &\geq \max_{\psi \in \mbcf{P}(\mbcf{Y}|\mbcf{Z})} \Pr \left(  \sum_{\mbf{y}} \psi(\mbf{y}|\mbf{Z}) \varphi(\mcf{M}-\{M\}|\mbf{y},K) \geq 2^{-n\alpha}  \right)  \label{eq:lem:c:1:codeass:-1}\\
&\geq  \Pr \left(  \sum_{\mbf{y},k,t} p(\mbf{y}|k,t) p(k,t|\mbf{Z})  \varphi(\mcf{M}-\{M\}|\mbf{y},K) \geq 2^{-n\alpha}  \right)  \label{eq:lem:c:1:codeass} \\
&\geq  \Pr \left(  \sum_{\mbf{y}:(\mbf{y},k,t) \in \mcf{D}^+} p(\mbf{y}|K,T) p(K,T|\mbf{Z})  \varphi(\mcf{M}-\{M\}|\mbf{y},K) \geq 2^{-n\alpha}  \right) \\
&=  \sum_{\mbf{z},m,k,t} \idc{ p(k,t|\mbf{z}) \tau(m,k,t)  \geq 2^{-n\alpha}  } p(\mbf{z},m,k,t) \label{eq:lem:c:1:codeass2}
\end{align} 
since $\sum_{k,t} p_{\mbf{Y}|K,T}(\cdot|k,t) p_{K,T|\mbf{Z}}(k,t|\cdot ) \in \mbcf{P}(\mbcf{Y}|\mbcf{Z})$ and all summands inside the probability term are positive.
Furthermore, from Equation~\eqref{eq:lem:c:1:codeass2}, it follows that
\begin{align}
\delta + \sqrt{\nu_n} + 2^{-n\lambda_n}  &\geq  \sum_{\mbf{z},t} \idc{ 2^{-\mathbb{H}(K|\mbf{Z},T=t)-3n\lambda_n} p(t|\mbf{z}) (1-\sqrt{\nu_n})  \geq 2^{-n\alpha}  }  p(\mbf{z},t) \label{eq:lem:c:1:codeimp2}
\end{align}
since 
$$p(k,t|\mbf{z}) = \frac{p(\mbf{z}|k,t) p(k|t)}{p(\mbf{z}|t)} p(t|\mbf{z}) \geq 2^{-\mathbb{H}(K|\mbf{Z},T=t) -3n\lambda_n} p(t|\mbf{z})$$
for all $(\mbf{z},m,k,t) \in \mcf{D}^+$ implies
\begin{align*} 
\idc{ p(k,t|\mbf{z}) \tau(m,k,t)  \geq 2^{-n\alpha}  } &\geq \idc{ 2^{-\mathbb{H}(K|\mbf{Z},T=t)-3n \lambda_n} p(t|\mbf{z}) (1-\sqrt{\nu_n})  \geq 2^{-n\alpha}  } \\
&\quad - \idc{ \tau(m,k,t) < 1-\sqrt{\nu_n}} - \idc{(\mbf{z},m,k,t) \notin \mcf{D}^+} 
\end{align*} 
and the sum of $\idc{ \tau(m,k,t) < 1-\sqrt{\nu_n}}$ terms can be bounded using Lemma~\ref{lem:pre:c:2}, while the sum of $\idc{(\mbf{z},m,k,t) \notin \mcf{D}^+}$ terms is bounded by the fact that $\Pr\left( (\mbf{Z},K,M,T) \notin \mcf{D}^+ \right) \leq 2^{-n\lambda_n} $.

Now, Equation~\eqref{eq:lem:c:eq:2} can be proved using basic information inequalities as follows:
\begin{align}
&\mathbb{H}(K|\mbf{Z}) + \mathbb{H}(T|\mbf{Z},K) + 3n\lambda_n - \logt( 1 - \sqrt{\nu_n}) \notag \\
&\quad= \sum_{\mbf{z},t} \left[ \mathbb{H}(K|\mbf{Z},T=t) + 3n\lambda_n -\logt p(t|\mbf{z}) - \logt(1-\sqrt{\nu_n}) \right] p(\mbf{z},t) \\
&\quad \geq n \alpha \sum_{\mbf{z},t}  \idc{ 2^{-\mathbb{H}(K|\mbf{Z},T=t)-3n\lambda_n} p(t|\mbf{z}) (1-\sqrt{\nu_n})  < 2^{-n\alpha}  }    p(\mbf{z},t) \\
&\quad\geq n\alpha( 1 - \delta - \sqrt{\nu_n} - 2^{-n\lambda_n})  \label{eq:lem:c:1:fin}
\end{align}
where~\eqref{eq:lem:c:1:fin} follows from Equation~\eqref{eq:lem:c:1:codeimp2}.
Hence, 
\begin{align}
n^{-1}\mathbb{H}(K|\mbf{Z}) + \zeta(\delta,n)
&\geq \alpha \label{eq:lem:c:1:finnish}
\end{align}
where
$$\zeta(\delta,n) = \frac{ (\delta+ \sqrt{\nu_n} + 2^{-n\lambda_n}) \kappa +  4\lambda_n  - n^{-1} \logt (1-\sqrt{\nu_n})}{1 - \delta - \sqrt{\nu_n} -2^{-n\lambda_n}},$$
since $\mathbb{H}(K|\mbf{Z}) \leq \mathbb{H}(K) =  n\kappa $ and $\mathbb{H}(T|\mbf{Z},K) \leq \logt|\mcf{T}|\leq n \lambda_n.$
Equation~\eqref{eq:lem:c:1:finnish} proves Equation~\eqref{eq:lem:c:eq:2} since $\lim_{\subalign{\delta &\rightarrow 0^+\\ n&\rightarrow \infty}} \zeta(\delta,n) = 0  .$

\end{IEEEproof}

\section{Proof of Theorem~\ref{thm:WAAR}}\label{app:WAAR_proof}

The proof of Theorem~\ref{thm:WAAR} is divided into three parts.
In Appendix~\ref{app:WAAR:direct}, it will be shown that if positive real numbers satisfy 
\begin{align*}
r + \alpha &\leq \mathbb{I}(Y;U,W)  \\
2 \alpha - \kappa  &\leq   \mathbb{I}(Y;U|W) -  \mathbb{I}(Z;U|W)   \\
\alpha - \kappa &\leq 0 
\end{align*}
for some random variables $X,U,W$ such that $W \markov U \markov X \markov (Y,Z)$, then $(r,\alpha, \kappa) \in \mcf{C}_{\mathrm{TA}}(p_{Y|X},p_{Z|X})$.
Next, in Appendix~\ref{app:WAAR:converse}, it will be shown that if $(r,\alpha, \kappa) \in \mcf{C}_{\mathrm{TA}}(p_{Y|X},p_{Z|X})$, then there exists $X,U,W$ such that $W \markov U \markov X \markov (Y,Z)$ and
\begin{align*}
r + \alpha &\leq \mathbb{I}(Y;U,W)  \\
2 \alpha - \kappa  &\leq   \mathbb{I}(Y;U|W) -  \mathbb{I}(Z;U|W)   \\
\alpha - \kappa &\leq 0.
\end{align*}
Finally, we shall show in Appendix~\ref{app:WAAR:card}, that restricting auxiliary random variables $U$ and $W$ so that $|\mcf{U}| \leq (|\mcf{X}|+1) (|\mcf{X}|+2)$ and $|\mcf{W}| \leq |\mcf{X}|+2$ does not reduce the established region.

\begin{IEEEproof}

\subsection{Direct for Theorem~\ref{thm:WAAR}}\label{app:WAAR:direct}

The set of all positive $(r,\alpha,\kappa)$ that satisfy
\begin{align}
r &= r' - \beta \label{eq:WAAR:d:a}\\
\alpha &= \alpha' + \beta \\
\kappa &= \kappa' + 2\beta \\
\beta &< r' \\
\beta &\geq 0\\
r'&\geq 0\\
\alpha' &\geq 0\\
\kappa' &\geq 0 \\
r' + \alpha' &< \mathbb{I}(Y;U,W) \\
\alpha' &< \mathbb{I}(Y;U|W) - \mathbb{I}(Z;U|W) \\
\alpha' - \kappa' &< 0 \label{eq:WAAR:d:b}
\end{align}
for some random variables $U,W$ where $W \markov U \markov X \markov (Y,Z)$ and real number $r',\alpha',\kappa',\beta$ is achievable by the combination of Theorem~\ref{thm:lai} and Theorem~\ref{lem:-t}.
Here, $(r',\alpha',\kappa')$ correspond to the points achievable by Theorem~\ref{thm:lai}, while $r,~\alpha,~\kappa,$ and $\beta$ correspond to the regions that can be obtained by applying Theorem~\ref{lem:-t} to Theorem~\ref{thm:lai}.
Applying Fourier-Motzkin elimination to remove $r',\alpha',\kappa'$ and $\beta$ from Equations~\eqref{eq:WAAR:d:a}--\eqref{eq:WAAR:d:b} proves that if 
\begin{align}
r + \alpha &< \mathbb{I}(Y;U,W) \label{eq:WAAR:d:1}\\
2\alpha - \kappa  &< \mathbb{I}(Y;U|W) - \mathbb{I}(Z;U|W) \\
\alpha - \kappa &< 0 \label{eq:WAAR:d:3},
\end{align}
then $(r,\alpha,\kappa) \in \mcf{C}_{\mathrm{TA}}(p_{Y|X},p_{Z|X}).$

\subsection{Converse for Theorem~\ref{thm:WAAR}}\label{app:WAAR:converse}

In order for $(r,\alpha,\kappa)$ to be achievable for DM-AIC$(p_{Y|X},p_{Z|X})$, there must exist a sequence of $(r_n,\alpha_n,\kappa_n,\delta_n,n)$-TA codes, for $n = \{1,2,\dots\}$, such that 
\[
\lim_{n \rightarrow \infty}   \abs{(r_n,\alpha_n,\kappa_n,\delta_n) - (r,\alpha,\kappa,0)} = 0.
\]
But,
\begin{align}
r_n + \alpha_n  &\leq n^{-1}\mathbb{I}(\mbf{Y} ;M,K) +\zeta(\delta_n,n) + n^{-1} + \delta_n \logt |\mcf{Y}|  \label{eq:conv:1b:1}\\
2\alpha_n -\kappa_n   &\leq  n^{-1}\left[ \mathbb{I}(\mbf{Y};M,K) - \mathbb{I}(\mbf{Z};M,K) \right] + 2\zeta(\delta_n,n) + n^{-1} + \delta_n \logt|\mcf{Y}| \label{eq:conv:1b:2}\\
\alpha_n -\kappa_n &\leq 0 \label{eq:conv:1b:3}
\end{align}
must hold for a given $(r_n,\alpha_n,\kappa_n,\delta_n,n)$-TA code.
Indeed, to prove Equations~\eqref{eq:conv:1b:1}--\eqref{eq:conv:1b:3}, first observe the following inequalities for a $(r_n,\alpha_n,\kappa_n,\delta_n,n)$-TA code:
\begin{align}
r_n  &\leq n^{-1}\mathbb{I}(\mbf{Y} ;M|K) + n^{-1} + \delta_n \logt |\mcf{Y}|  \label{eq:conv:1:1}\\
0  &\leq n^{-1}[ \mathbb{I}(\mbf{Y};M|K) -  \mathbb{I}(\mbf{Z};M|K) ] + n^{-1} + \delta_n \logt|\mcf{Y}| \label{eq:conv:1:1b}\\
\alpha_n    &\leq  n^{-1} \mathbb{I}(\mbf{Y};K) + \zeta(\delta_n,n)  \label{eq:conv:1:2}\\
\alpha_n  &\leq n^{-1}\mathbb{H}(K|\mbf{Z}) + \zeta(\delta_n,n) \label{eq:conv:1:3}\\
\kappa  &= n^{-1}\mathbb{H}(K)  \label{eq:conv:1:3b}
\end{align}
where \eqref{eq:conv:1:1} and~\eqref{eq:conv:1:1b} are because (from Fano's inequality and the data processing inequality)
$$\mathbb{H}(M|K) = \mathbb{I}(\mbf{Y};M|K) + \mathbb{H}(M|\mbf{Y},K) \leq \mathbb{I}(\mbf{Y};M|K) + 1+ n\delta_n \logt |\mcf{Y}|  ,$$
and further for~\eqref{eq:conv:1:1} because $nr_n = \mathbb{H}(M|K)$ and for~\eqref{eq:conv:1:1b} because $\mathbb{I}(\mbf{Z};M|K) \leq \mathbb{H}(M|K)$; next~\eqref{eq:conv:1:2} and~\eqref{eq:conv:1:3} are due to Lemma~\ref{lem:c:2}; and finally~\eqref{eq:conv:1:3b} is because $K$ is uniform over $\{1,\dots,2^{n\kappa}\}.$
Equations~\eqref{eq:conv:1b:1}--\eqref{eq:conv:1b:3} can be derived from linear combinations of~\eqref{eq:conv:1:1}--\eqref{eq:conv:1:3b}.

Now, Equations~\eqref{eq:conv:1b:1}--\eqref{eq:conv:1b:3} also dictate (as we will show in later in the proof) that there exists RVS $X,U,W$, such that  $W\markov U \markov X \markov (Y,Z),$ and  
\begin{align}
r_n + \alpha_n  &\leq \mathbb{I}(Y;U,W) +\zeta(\delta_n,n) + n^{-1} + \delta_n \logt |\mcf{Y}|  \label{eq:conv:2b:1}\\
2\alpha_n -\kappa_n   &\leq   \mathbb{I}(Y;U|W) - \mathbb{I}(Z;U|W) + 2\zeta(\delta_n,n) + n^{-1} + \delta_n \logt|\mcf{Y}| \label{eq:conv:2b:2}\\
\alpha_n -\kappa_n &\leq 0 \label{eq:conv:2b:3}.
\end{align}
Furthermore, we can take without loss of generality $|\mcf{U}|\leq (|\mcf{X}|+2)(|\mcf{X}|+1)$ and $|\mcf{W}|\leq |\mcf{X}|+2$ as shown in Appendix~\ref{app:WAAR:card}.
Thus, each triple $(r,\alpha,\kappa) \in \mcf{C}_{\mcf{TA}}(p_{Y|X},p_{Z|X})$ can be described by a limit point of the set of $(r',\alpha',\kappa')$ for which there exist $X,U,W$, where $|\mcf{U}|\leq (|\mcf{X}|+2)(|\mcf{X}|+1)$ and $|\mcf{W}|\leq |\mcf{X}|+2$, such that $W\markov U \markov X \markov (Y,Z)$ and
\begin{align}
r' + \alpha'  &\leq \mathbb{I}(Y;U,W)  \label{eq:conv:3b:1'}\\
2\alpha' -\kappa'   &\leq   \mathbb{I}(Y;U|W) - \mathbb{I}(Z;U|W)  \label{eq:conv:3b:2'}\\
\alpha' -\kappa' &\leq 0 \label{eq:conv:3b:3'}
\end{align}
since $\lim_{n\rightarrow \infty} \delta_n = 0 $ and $\lim_{n\rightarrow \infty} \zeta(\delta_n,n)= 0 .$
But, the set of $(r',\alpha',\kappa')$ that satisfy Equations~\eqref{eq:conv:3b:1'}--\eqref{eq:conv:3b:3'} is a closed set by~\cite[Theorem~4.15]{rudin1964principles} since the set of all probability mass functions of $X,U,W$, where $|\mcf{U}|\leq (|\mcf{X}|+2)(|\mcf{X}|+1)$, and $|\mcf{W}|\leq |\mcf{X}|+2$, and $W\markov U \markov X \markov (Y,Z)$, is itself a compact set.
Hence, it follows that if $(r,\alpha,\kappa) \in \mcf{C}_{\mcf{TA}}(p_{Y|X},p_{Z|X})$, then there exists $X,U,W$, where $|\mcf{U}|\leq (|\mcf{X}|+2)(|\mcf{X}|+1)$ and $|\mcf{W}|\leq |\mcf{X}|+1$, such that $W \markov U \markov X \markov (Y,Z)$ and
\begin{align}
r + \alpha  &\leq \mathbb{I}(Y;U,W)  \label{eq:conv:3b:1}\\
2\alpha -\kappa   &\leq   \mathbb{I}(Y;U|W) - \mathbb{I}(Z;U|W)  \label{eq:conv:3b:2}\\
\alpha -\kappa &\leq 0 \label{eq:conv:3b:3}.
\end{align}

We now return to proving Equations~\eqref{eq:conv:2b:1}--\eqref{eq:conv:2b:3}.
This can be done via a trick from the proof of~\cite[Lemma~15.7]{CK}, in which  
for $\mbf{Y} = (Y_1,\dots,Y_n)$ and $\mbf{Z} = (Z_1,\dots,Z_n)$ it is shown that 
\begin{align}
n^{-1}[ \mathbb{I}(\mbf{Y};M,K) - \mathbb{I}(\mbf{Z};M,K)]  &= n^{-1}[ \mathbb{I}(Y_1;M,K|Z_2^n) - \mathbb{I}(Z_1;M,K|Z_2^n) ] \notag \\
&\quad + n^{-1}\sum_{i=2}^n [ \mathbb{I}(Y_2^n;M,K|Y_1) - \mathbb{I}(Z_2^n;M,K|Y_1) ]  \\
&= \sum_{i=1}^{n} n^{-1}[ \mathbb{I}(Y_i;M,K|Y_1^{i-1},Z_{i+1}^n) - \mathbb{I}(Z_i;M,K|Y_1^{i-1},Z_{i+1}^n) ] \\
&= \mathbb{I}( Y_J;M,K|W) - \mathbb{I}( Z_J;M,K|W) \label{eq:conv:up1}
\end{align}
where $J$ is uniformly distributed over $\{1,\dots,n\}$, and $W \defn (Y_{1}^{J-1},Z_{J+1}^n,J)$.
Now, clearly, $$(W,M,K) \markov X_J \markov Y_J,$$ $p_{Y_J|X_J} = p_{Y|X}$, and 
\begin{align}
n^{-1} \mathbb{I}(\mbf{Y};M,K) &= \sum_{i=1}^n n^{-1} \mathbb{I}(Y_i;M,K|Y_1^{i-1}) \leq \sum_{i=1}^{n} n^{-1}\mathbb{I}(Y_i;M,K,Y_1^{i-1},Z_{i+1}^n) = \mathbb{I}( Y_J; M,K,W|J) \\
&\leq \mathbb{I}(Y_J;M,K,W). \label{eq:conv:up2}
\end{align}
Combining Equations~\eqref{eq:conv:up1} and~\eqref{eq:conv:up2} with Equations~\eqref{eq:conv:1b:1}--\eqref{eq:conv:1b:3} and setting $U = (W,M,K)$ yields~\eqref{eq:conv:2b:1}--\eqref{eq:conv:2b:3}.

\subsection{Auxiliary random variable cardinalities}\label{app:WAAR:card}

Finally, we now return to prove that $|\mcf{U}|\leq(|\mcf{X}|+2)(|\mcf{X}|+1)$ and $|\mcf{W}|\leq |\mcf{X}|+2$ in Equations~\eqref{eq:conv:2b:1}--\eqref{eq:conv:2b:3}. 
This can be done via the Fenchel--Eggleston--Carath{\'e}odory theorem (see, for example,~\cite[Appendix~A]{GNIT} or~\cite[Lemma~15.6]{CK}).
For completeness, we will prove the bounds using a restricted version of a support lemma from~\cite[Appendix~C]{GNIT}.
Note, we enter here a restricted version of the lemma, because the general $\mcf{U}$ and $\mcf{W}$ from Equations~\eqref{eq:conv:2b:1}--\eqref{eq:conv:2b:3} has a finite support set, and thus we have no need to discuss continuous distributions or differential entropy.  
\begin{lemma} \textbf{(\cite[Appendix~C]{GNIT})} \label{lem:sup} Let $\mcf{X}$ and $\mcf{U}$ be finite sets. Let $\mcf{Q}$ be a connected compact subset of pmfs of $\mcf{X}$ and $p_{X|U=u} \in \mcf{Q}$ for each $u \in \mcf{U}$.  
Suppose that  $g_{j}(\pi)$, $j = 1,\dots,d$, are real valued continuous functions of $\pi \in \mcf{Q}.$ Then for every $U \sim p_{U}(u)$ defined on $\mcf{U},$ there exists a random variable $U'\sim p_{U'}(u')$ with $|\mcf{U'}| \leq d$ and $p_{X|U'=u'} \in \mcf{Q}$, for each $u' \in \mcf{U}',$ such that for $j = 1,\dots , d,$
$$ \sum_{u \in \mcf{U}} g_{j}(p_{X|U=u}) p_{U}(u) = \sum_{u' \in \mcf{U}'} g_{j}(p_{X|U'=u'}) p_{U'}(u').$$
\end{lemma}

With Lemma~\ref{lem:sup}, the goal is to find $U'$ and $W'$ such that $|\mcf{U}'| \leq (|\mcf{X}|+1)(|\mcf{X}|+2)$ and $|\mcf{W}'|\leq |\mcf{X}|+2$, as well as $W'\markov U' \markov X \markov (Y,Z)$ and 
\begin{align}
\mathbb{I}(Y;U,W) &= \mathbb{I}(Y;U',W') \label{eq:card:fit1}\\
\mathbb{I}(Y;U|W) - \mathbb{I}(Z;U|W)  &=   \mathbb{I}(Y;U'|W') - \mathbb{I}(Z;U'|W'). \label{eq:card:fit2}
\end{align}
Doing so shows that we may restrict the cardinalities of $U$ and $W$.

First, to replace $W\sim p_{W}(w)$ with $W'\sim p_{W'}(w)$, observe that $p(x)$ (for each $x$), $\mathbb{H}(Y|W),$ $\mathbb{I}(Y;U|W),$ and $\mathbb{I}(Z;U|W)$ can be written as
\begin{align}
p(x) &= \sum_{w\in \mcf{W}} p(x|w) p_{W}(w), \quad   \forall x \in \mcf{X} \\
\mathbb{H}(Y|W) &= \sum_{w\in \mcf{W}} \mathbb{H}(Y|W=w) p_{W}(w) \\
\mathbb{I}(Y;U|W) &=  \sum_{w\in \mcf{W}}\mathbb{I}(Y;U|W=w) p_{W}(w) \\
\mathbb{I}(Z;U|W) &=  \sum_{w\in \mcf{W}}\mathbb{I}(Z;U|W=w)p_{W}(w) 
\end{align}
and that $p(x|w)$ (for each $x$), $\mathbb{H}(Y|W=w)$, $\mathbb{I}(Y;U|W=w)$, and $\mathbb{I}(Z;U|W=w)$ are each a continuous function of distribution $p_{U|W=w}.$ 
Thus, there exists a $\mcf{W}'$ and $W'$ such that $|\mcf{W}|' \leq |\mcf{X}|+2$ (note that fixing $p(x)$ for $x \in \{1,\dots, |\mcf{X}|-1\}$ also fixes $p(|\mcf{X}|)$) and 
\begin{align}
 \sum_{w\in \mcf{W}} p(x|w) p_{W}(w)  &= \sum_{w'\in \mcf{W}' } p(x|w') p_{W'}(w'), \quad   \forall x \in \mcf{X} \label{eq:card:y}\\
\mathbb{H}(Y|W) &= \mathbb{H}(Y|W') \label{eq:card:y+1}\\
\mathbb{I}(Y;U|W) &=  \mathbb{I}(Y;U'|W') \label{eq:card:y2-1}  \\
\mathbb{I}(Z;U|W) &= \mathbb{I}(Z;U'|W'), \label{eq:card:y2}
\end{align}
where $U'|\{W'=w'\} $ is $U|\{W=w'\},$ by Lemma~\ref{lem:sup}.
Furthermore, 
\begin{equation}\label{eq:card:ra1}
\mathbb{I}(Y;U,W) = \mathbb{I}(Y;U|W) + \mathbb{H}(Y) - \mathbb{H}(Y|W) = \mathbb{I}(Y;U',W')
\end{equation}
by Equations\footnote{Equation~\eqref{eq:card:y} implies that $\mathbb{H}(Y)$ remains unchanged when attaching $W$ or $W'.$}~\eqref{eq:card:y}--\eqref{eq:card:y2-1}
while 
\begin{equation}\label{eq:card:diff1}
\mathbb{I}(Y;U|W) - \mathbb{I}(Z;U|W)  = \mathbb{I}(Y;U'|W') - \mathbb{I}(Z;U'|W') 
\end{equation}
by Equations~\eqref{eq:card:y2-1} and~\eqref{eq:card:y2}.
Equations~\eqref{eq:card:ra1} and~\eqref{eq:card:diff1} demonstrate that we may assume that $|\mcf{W}| \leq |\mcf{X}|+2$ in Equations~\eqref{eq:card:fit1} and~\eqref{eq:card:fit2}.

So, let us assume that $|\mcf{W}| \leq |\mcf{X}|+2$, but that $\mcf{U}$ is arbitrary.  
This time, in order to replace $U$, where $ U|\{W=w\} \sim  \frac{\dx}{\dx u} p_{U|W=w}$, with $U'$ where $U'|\{W=w\} \sim p_{U'|W}(u|w)$, observe that we only need to conserve $p(x|w)$ (for each $x\in \{1,\dots,|\mcf{X}|-1\}$ and $w$), $\mathbb{H}(Y|U,W=w)$ (for each $w$), and $\mathbb{H}(Z|U,W=w)$ (for each $w$).
To this end, as before, each of these $(|\mcf{X}|+1)|\mcf{W}|$ equations can be written as the average of a continuous function of $p_{X|U=u}.$
Therefore, there exists a $U'$ and $\mcf{U}'$, where $|\mcf{U}'| \leq (|\mcf{X}|+1)|\mcf{W}|\leq (|\mcf{X}|+1)(|\mcf{X}|+2)$, such that 
\begin{equation}\label{eq:card:ra2}
\mathbb{I}(Y;U,W) =  \mathbb{I}(Y;U',W)
\end{equation}
and
\begin{equation}\label{eq:card:diff2}
\mathbb{I}(Y;U|W)-\mathbb{I}(Z;U|W) = \mathbb{I}(Y;U'|W)-\mathbb{I}(Z;U'|W) 
\end{equation}
by Lemma~\ref{lem:sup}.

\end{IEEEproof}

\bibliographystyle{ieeetr}
\bibliography{this}

\end{document}